\documentclass[10pt,a4paper]{IEEEtran}
\usepackage{amsthm}
\usepackage{epsfig,psfrag}
\usepackage{threeparttable}
\usepackage{amsmath}
\usepackage{mathtools}
\usepackage{amssymb}
\usepackage{makecell}
\usepackage{here}
\usepackage[para]{footmisc}
\usepackage{bbm}
\usepackage{mathrsfs}
\usepackage{booktabs}
\usepackage{dsfont}
\usepackage{multirow}

\newcounter{lawc}
\newtheorem{theorem}{Theorem}

\newtheorem{proposition}[theorem]{Proposition}

\newtheorem{define}[theorem]{Definition}
\newtheorem{example}[theorem]{Example}

\newtheorem{result}[theorem]{Result}
\newtheorem{fact}[theorem]{Fact}

\newcommand{\menlem}{\hfill \ensuremath{\blacktriangle}}
\newcommand{\mendprinciple}{\hfill \ensuremath{\therefore}}
\newcommand{\mendprop}{\hfill \ensuremath{\triangledown}}

\DeclareMathOperator*{\col}{col}
\DeclareMathOperator{\Sq}{Sq}
\DeclareMathOperator{\sign}{sign}
\DeclareMathOperator*{\diag}{diag}
\DeclareMathOperator{\eps}{\varepsilon}
\DeclareMathOperator{\RTT}{RTT}

\newenvironment{law}
{\noindent\textbf{Law of Conservation of Information:}\\\begin{itshape}}
{\mendprinciple\end{itshape}}

\title{The conservation of information, towards an axiomatized modular modeling approach to congestion control}

\author{C.~Briat, E.A. Yavuz\\ H. Hjalmarsson, K.H.~Johansson, U.T. J\"{o}nsson, G.~Karlsson, and H.~Sandberg\thanks{This work has been supported by ACCESS Linnaeus Center, KTH, Stockholm, Sweden. http://www.access.kth.se/, e-mail: \{cbriat,emreya,hakan.hjalmarsson,gk,kallej,hsan\}@kth.se}}

\begin{document}
\maketitle

%


\begin{abstract}
We derive a modular fluid-flow network congestion control model
based on a law of fundamental nature in networks: the conservation of information. Network elements such as queues, users, and transmission channels and
network performance indicators like sending/acknowledgement rates
and delays are mathematically modelled by applying this law
locally. Our contributions are twofold. First, we introduce a
modular metamodel that is sufficiently generic to represent any
network topology. The proposed model is composed of building
blocks that implement mechanisms ignored by the existing ones,
which can be recovered from exact reduction or approximation of
this new model. Second, we provide a novel classification of
previously proposed models in the literature and show that they
are often not capable of capturing the transient behavior of the
network precisely. Numerical results obtained from packet-level
simulations demonstrate the accuracy of the proposed model.

\end{abstract}


\section{Introductory discussions}

\subsection{The congestion control problem}

The congestion problem \cite{Jacobson:88,Srikant:04} is inherent to communication networks where
capacity of supporting infrastructure that relays information is
small compared to user demand. Congestion is responsible for delay
and data loss, which compromise the efficiency of the overall
network. Controlling congestion is hence an important problem for
which several algorithms have been developed. They mainly rely on
the concept of \emph{congestion window}\footnote{the number of
desired outstanding packets}, which is adapted according to a
\emph{congestion measure}.
%
%
According to the type of congestion measure, two classes of
congestion control algorithms may be identified \cite{Low:02,Srikant:04}. The first and
oldest class is \emph{loss-based}, meaning that the congestion
measure is the packet-loss information. This class is easy to
implement but leads to a quite rough control since the protocol
detects the network congestion only after provoking it. In order
to control congestion more smoothly and prevent data loss,
\emph{delay-based} algorithms, using for instance the
\emph{Round-Trip Time} (RTT) information as the congestion
measure, may be considered instead. They are however more
difficult to implement due to the possible unavailability of
certain necessary measures, such as \emph{queuing delays}.

The main difficulty in congestion control lies in the fact that, basically, the hosts ignore almost everything about the network: the routes and their capacity, the numbers of routers (hops), the number of users, etc. Hence, protocol designers face the problem of controlling a very large and complex system with actually very little information. 

When designing a protocol, \emph{stability} of the network is
certainly the most important constraint. Performance criteria can
be additionally considered in order to optimize the network
behavior. For instance, we may want to use all available bandwidth
(\emph{efficiency}), share it equally between users
(\emph{fairness}), and/or be tolerant with respect to unregulated
traffic and other protocols (\emph{cross-traffic adaptation}).

In order to observe/predict the network behavior and validate a
protocol, simulations and experiments must usually be conducted.
NS-2 is a widely accepted open-source event-based simulator
dedicated to this purpose. However, as any other simulator, it
does not permit to analyze the stability of a network
theoretically. Hence constructing mathematical models for networks
may play an important role in network analysis and protocol design
since they potentially allow for a theoretical analysis and an
equation-based design of new protocols.

\subsection{Models, approximations and accuracy}

Modeling is now ubiquitous. The key idea is to start from a system
and arrive at an abstract representation of it, such as one given
in terms of a set of mathematical equations. It is not always
necessary that a model represents all characteristics of a system
but only a subset of interest: e.g. a molecular-level model can be
irrelevant to portray a river. This gave rise to fluid-mechanics
which, although being an idealization of the reality, yields very
accurate predictions. A similar idealization has been shown to be
very useful for the congestion control problem through the
consideration of \emph{fluid-flow models} \cite{Mitra:88}.

\subsection{Metamodels and network models}

A paragon of network modeling is undoubtedly used in the field of electrical engineering. It is easy to identify the reasons for the success of the theoretical framework:
\begin{enumerate}
  \item Only two universal concepts: current and voltage, governed by simple laws (Kirchhoff's laws)
  \item Local description of the elements in terms of these variables and additional local concepts (e.g. resistance, etc.)
  \item Easy transcription of the electrical network into a topologically identical diagram, and vice-versa.
  \item New models corresponding to new devices may be freely added without compromising the existing ones.
  \item Model predictions fit very well to reality.
  \item Systematic way of analysis by hand calculations or simulators.
\end{enumerate}
A very important feature is that the principles of modeling an
electrical network is independent of its topology and elements.
This is achieved thanks to the structure of the paradigm that we
refer from now on as a \emph{metamodel}, which is a model that
consists of a set of frames, rules, constraints, submodels and
theories applicable and useful for modeling a predefined class of
problems. In the case of electrical networks, the metamodel
consists of  the concepts of current and voltage, the Kirchhoff's
laws and the local models of electrical elements (resistor,
capacitor, transistor, etc.) as well as all the related
mathematical tools. Since networks (like communication networks,
electrical networks, transportation networks and even social
networks) consist of interconnections of several elements, it
turns out that metamodels are then very suitable for describing
them since they also consist of interconnection of concepts, rules
and submodels. Hence metamodels provide, in essence, an elegant
scalable and modular way for modeling networks.

\subsection{Motivations and contributions}

The main motivation of this work is to give a clear picture of
congestion control modeling problem through derivation of a
metamodel having solid mathematical foundations. We introduce a modular metamodel that would lead to an interesting step forward towards a generic way of providing models for communication networks. This metamodel should then satisfy the
additional constraints on independence of network topology
(scalability) and elements (modularity). It should also provide
accurate predictions along with simple graph representation. An
underlying difficulty is the presence of several phenomena at
different levels: decision to send a packet, transmission of
packets on transmission channels, storage of packets in queues and
time-varying waiting-time (queuing delays), congestion window size
adaptation, etc. Finding a unified way for representing all these
critical phenomena is challenging.

The proposed metamodel is based on a single concept of information conservation, from which models for the network constituents (i.e. transmission channels, queues and users) are obtained. This allows to derive new models for network elements, obtain mathematical proofs for unproved/claimed existing ones, and invalidate some of them. All important variables of the network (sending rates, ACK rates, queue size, etc) are described by explicit formulas, hence computable. Using the proposed metamodel, describing a given topology is immediate, and performed by simply plugging the models together, so as the actual topology is transcribed into a graph having the different network elements located on the edges, as in electrical engineering. It is also proved that existing sending rate models are either approximation of the proposed sending rate model, or even exact provided that the network topology satisfies certain structural conditions. Simulations and comparisons with existing works tend to suggest the relevance, reliability and accuracy of the proposed metamodel. A non-exhaustive summary of related works on congestion control modeling is finally made in order to compare congestion control models according to important properties and criteria\footnote{The authors are aware of the fact that all the models are certainly not listed. This is however a first attempt and any suggestion from the reviewers to improve/correct this comparison is welcome.}.

The outline of the paper is as follows: Section \ref{sec:netgraph}
introduces the particular network graph representation considered
in the paper. Using continuous-time models, such as
fluid-flow models, to describe networks is justified and concepts of
universal clock, local discrete-time network element clock, and
clock-coupling are defined in section \ref{sec:fluidflow}. Section
\ref{sec:conservation} presents the conservation law of
information and we derive the transmission channel model in
section \ref{sec:transmed}. In section \ref{sec:buffer}, we
develop the model for the FIFO buffer network element. The user
model is given in section \ref{sec:user} and we summarize the
obtained results in a compact form in section \ref{sec:general}.
In section \ref{sec:SBMUtopology}, we consider a network with
single buffer/multiple-user topology to implement the proposed
model. We validate our model in section \ref{sec:validation} and
related work is given in section \ref{sec:related}. Section
\ref{sec:conclusion} concludes the paper.

\section{Networks and Graphs}\label{sec:netgraph}



It is convenient to introduce here the particular network graph representation considered in the paper. It is different from the traditional ones \cite{Johari:00,Hollot:01,Vinnicombe:02,Paganini:03} since it places all network elements on graph edges, leaving nodes with the role of connecting points, as in electrical circuits. Four types of nodes are distinguished: the input nodes $u_i^-$, $b_j^-$ and output nodes $u_i^+$, $b_j^+$ for user $u_i$ and buffer $b_j$ respectively. The superscripts have to be understood as a temporal order of reaction or causality: the data come (-) then leave (+). We will denote any edge $E$ of the graph by $\langle x,y\rangle$ where $x$ and $y$ are the input and output nodes respectively. Moreover, given any edge $E$, the input and output nodes are given by $\beta(E)$ and $\eps(E)$ respectively. 

\begin{figure}
  \centering
        \psfrag{be}[c][c]     {{{$\langle b^-,b^+\rangle$}}}
        \psfrag{bi-}[c][c]     {{{$b^-$}}}
        \psfrag{bi+}[c][c]     {{{$b^+$}}}
        \psfrag{ui-}[c][c]     {{{$u^-$}}}
        \psfrag{ui+}[c][c]     {{{$u^+$}}}
        \psfrag{ue}[c][c]     {{{$\langle u^-,u^+\rangle$}}}
        \psfrag{l1e}[c][c]     {{{$\langle u^+,b^-\rangle$}}}
        \psfrag{l2e}[c][c]     {{{$\langle b^+,u^-\rangle$}}}
        \psfrag{b}[c][c]     {{{$b$}}}
        \psfrag{l1}[c][c]     {{{$\ell_1$}}}
        \psfrag{l2}[c][c]     {{{$\ell_2$}}}
        \psfrag{u}[c][c]     {{{$u$}}}
  \includegraphics[width=0.25\textwidth]{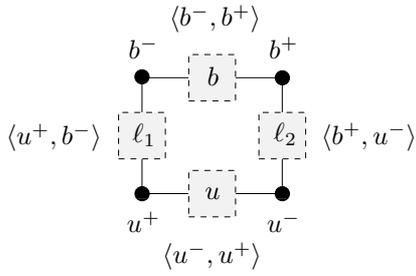}
  \caption{Example of graph with 4 edges: one user edge $u=\langle u^-,u^+\rangle$, one buffer edge $b=\langle b^-,b^+\rangle$ and two transmission edge $\ell_1=\langle u^+,b^-\rangle$, $\ell_2=\langle b^+,u^-\rangle$}\label{fig:graph}
\end{figure}

According to these definitions, a queue edge is always denoted by $\langle b_i^-,b_i^+\rangle$, a user edge by $\langle u_i^-,u_i^+\rangle$ and a transmission edge by $\langle b_i^+,u_j^-\rangle$, $\langle u_i^+,b_j^-\rangle$ or $\langle b_i^+,b_k^-\rangle$, $i\ne k$. This is illustrated in Fig. \ref{fig:graph}. We call a circuit $C_i$ the communication path of user $u_i$, that is the path connecting its output $u_i^+$ to its input $u_i^-$, i.e. $C_i=\langle u_i^+,u_i^-\rangle$. Note that in complex networks there exist several possible paths but only one of them, the one used for communication, is a circuit. In Fig. \ref{fig:graph}, the only possible circuit is given by $C=\langle u^+,b^-,b^+,u^-\rangle$.

\section{Fluid-flow idealization}\label{sec:fluidflow}


In an asynchronous network like the Internet, each element can be considered to have its own local discrete-time clock $\mathbb{T}_i\subset\mathbb{R}_+$ where $\mathbb{T}_i$ is countable, governing the rhythm of protocol decisions and packets transmission. In congestion control, the clocks beat with the rhythms of acknowledgment reception rates, which are influenced in turn by network congestion; this is referred to as \emph{ACK-clocking}\footnote{The term \emph{self-clocking} is also used but is less explicit.}. When several sources send data through the same buffer/path, a flow-coupling takes place leading then to clock-coupling. This clock-coupling arises at a very large-scale and distant sources having their clocks coupled cannot be considered to have independent behaviors. As a consequence, the sending rates and acknowledgment rates are hence intimately inter-dependent. Modeling this clock-coupling and the underlying phenomena is of incredible complexity since the number of clocks and their interactions grow very quickly with the network complexity, leading then to a very complicated structure for the interrelated local clocks $\mathbb{T}_i$, see for example \cite[Equations (3.7)]{Jacobsson:08}.


An idea to resolve this complex time-structural problem relies on the definition of a \emph{universal clock} $\mathbb{T}^u$ dictating a common time to the entire network. 
This leads us to the following fact:
\begin{fact}\label{ax:0}
  \emph{There exists an ideal universal clock $\mathbb{T}^u$ embedding any local clock $\mathbb{T}_i$, i.e. $\left(\bigcup_i\mathbb{T}_i\right)\subset\mathbb{T}^u$.}
  \mendprinciple
\end{fact}

A natural universal clock is given by $\bigcup_i\mathbb{T}_i$ and is a discrete-time clock. It however does not simplify too much the modeling problem since it is difficult to write recurrence relations for general network topologies \cite{Jacobsson:08}. Deriving a metamodel achieving scalability is then unlikely using such a universal clock. A however less natural universal clock $\mathbb{T}^u$ assimilated to a clock running
over positive real numbers continuously, i.e.
$\mathbb{T}^u\equiv\mathbb{R}_+$, is much more promising. This particular universal clock
indeed dramatically simplifies the modeling problem, and this motivates its consideration in this paper. Using such a time-scale, a metamodel
can be obtained, resulting then in a scalable solution in which
the network asynchrony is captured through appropriate expansions
and compressions of the time-space. Furthermore it enables the use
of well-established mathematical tools: real function analysis,
integration theory, dynamical systems, delay-differential
equations, etc. A conclusion is that continuous-time models may be
used to describe networks
\cite{Lindley:52,Mitra:88,Misra:00,Vinnicombe:00b,Hollot:01}: these are generally
referred to as \emph{fluid-flow models}, emphasizing the connection with continuum mechanics and more specifically with fluid-mechanics.

%
Within this framework, it is possible to provide a proper
definition for data flows.
\begin{define}
  Let us consider a point $x$ in an edge $E$ of the network and denote the number of packets having passed through point $x$ between $t_0\in\mathbb{T}^u$ and $t\in\mathbb{T}^u$, $t\ge t_0$, by $N_x(t,t_0)$. Then, the flow of data passing through point $x$ is defined as a function $\phi:E\times\mathbb{T}^u\to\mathbb{R}_+$ verifying:
\begin{equation}\label{eq:Nphi}
  N_x(t,t_0)=\int_{t_0}^t\phi(x,s)ds
\end{equation}
where the integral is a standard one, e.g. the Lebesgue integral.
\end{define}
Flows are defined in such a way rather than being the derivative of the number of packets, since the number of packets is non-differentiable, i.e. flows may contain dirac pulses, steps and so on. It is also interesting to note that since the the universal clock embeds all the local clocks, it is possible to recover discrete-time asynchronous models such as the one in \cite{Jacobsson:08} by setting flows to be trains of dirac pulses on $\bigcup_i\mathbb{T}_i$.

Using the notation defined in Section \ref{sec:netgraph}, we can build the flow vectors $\phi(x,t)$ using the '$\col$' operator:
\begin{equation}\label{eq:flow-+}
    \phi(x,t)=\col_{k=1}^{\eta(x)}\left[\phi_k(x,t)\right]
\end{equation}
where $x$ in any input and output node of the network elements, i.e. $x$ can be any $u_i^-$, $u_i^+$, $b_i^-$, $b_i^+$. The quantity $\eta(x)$ denotes the number of flows passing through node $x$. The concept of flows of data is hence very close to those of current in electrical engineering and flow of a liquid in fluid-mechanics.

\section{Conservation law of information}\label{sec:conservation}

The core of the metamodel is the conservation law of information
stated in this section. This law allows to improve the characterization of the elements of the network by notably clarifying their input/output relationship, enabling then a modular formalism. This conservation law follows from the remark that the
quantity of information\footnote{expressed in bits or packets.} is preserved in a communication network:
the data can either be in transit, lost or received. Assuming
lossless networks, it is possible to determine the total number of
packets in transit in any edge, simply by counting the number of
entering packets according to a simple rule. When applied to a specific element, this law allows to characterize the fact that the information is preserved from the input to the output.\\

\begin{law}
\emph{Given any edge $E$ of a network, then for all
$t\in\mathbb{T}^u$, there exists a time $t_0(t)\in\mathbb{T}^u$,
$t_0(t)\le t$ such that
\begin{equation}\label{eq:cl}
\begin{array}{lcl}
P_E(t)&:=&\int_{E}\phi(\theta,t)d\theta\\
&=&\int_{t_0(t)}^t\phi(\beta(E),s)ds\\
&=&N_{\beta(E)}(t,t_0(t))
\end{array}
\end{equation}
The integration over $E$ is an abstract integral which has to be understood as a flow spatial integration from $\beta(E)$ to $\eps(E)$, that is, the number of packets $P_E(t)$ in the edge $E=\langle\beta(E),\eps(E)\rangle$ at time $t$.}
\end{law}
The above result stated in quite abstract terms just says that the number of packets in transit in an edge at a certain time $t$ can be determined by counting the number of entering packets (i.e. integrating the input flow) over the interval $[t_0(t),t]$, the lower bound $t_0$ of the interval depending on the considered element, i.e. transmission channel, queue or user. A simple application of the law is given in the next section discussing transmission channels models.

The first benefit of this law is to show that we can interchangeably use a spatial or a temporal integral to calculate the quantity of information (number of packets) in an
edge. The temporal integral formulation is very convenient to work with since it requires the knowledge of the input flow only, rather than the flow value on the entire edge for the spatial integral formulation. This hence allows to
\emph{discretize the space dimension} by only considering flows at the
nodes, simplifying then the network representation and the modeling problem.

The second benefit lies in the fact that the temporal integral can be utilized to yield explicit solutions for the output flows of the different network elements. A very general result is given below:
\begin{proposition}\label{prop:ax1}
   The input flow $\phi(\beta(E),\cdot)$ and the output flow $\phi(\eps(E),\cdot)$ of edge $E$ verify
   \begin{equation}
  \phi(\eps(E),t)=t_0(t)^\prime\phi(\beta(E),t_0(t))
\end{equation}
where we assume that $t_0(t)$ is absolutely continuous and
$t_0(t)^\prime$ is the upper-right Dini derivative of $t_0$ at
$t$, i.e.
${t_0^\prime(t)=\limsup_{h\downarrow0}h^{-1}\left(t_0(t+h)-t_0(t)\right)}$.
\end{proposition}
\begin{proof}
Since $N_{\beta(E)}(t,t_0(t))$ is the current number of packets on
edge $E$ at time $t$, then differentiation with respect to time
provides the balance equation
  \begin{equation*}
  [N_{\beta(E)}(t,t_0(t))]^\prime=\phi(\beta(E),t)-t_0(t)^\prime\phi(\beta(E),t_0(t)).
\end{equation*}
Note also that a second valid balance equation is given by
  \begin{equation*}
  [N_{\beta(E)}(t,t_0(t))]^\prime=\phi(\beta(E),t)-\phi(\eps(E),t).
\end{equation*}
Identifying the right-hand side yields the result.
\end{proof}

The proposition given above plays a crucial role in the
metamodeling problem since it provides an explicit formula of the output flows. This output flow verifies the conservation of information from the input to the output of the edge $E$. By integrating the input and output over $[0,\infty)$ the very same value is obtained. This emphasizes that the output flow is defined in such a way that, as desired, it respects the natural property of conservation of information. Proposition \ref{prop:ax1} is used repeatedly in the paper in order to provide accurate and explicit models for transmission channels, queues and users. Applications are given in Sections \ref{sec:transmed}, \ref{sec:outputflow} and \ref{sec:ackflow}.

\section{Transmission Channel Model With Constant Propagation Delay}\label{sec:transmed}

%


\begin{figure}
  \centering
        \psfrag{d}[c][c]     {\small{Constant propagation delay $T$}}
        \psfrag{ch}[c][c]     {\small{\shortstack[l]{Transmission channel\\$\qquad\quad \langle x,y\rangle$}}}
        \psfrag{ip}[c][c]     {\small{$\phi(x,t)$}}
        \psfrag{op}[c][c]     {\small{\shortstack{$\phi(y,t)$\\$=$\\$\phi(x,t-T)$}}}
  \includegraphics[width=0.4\textwidth]{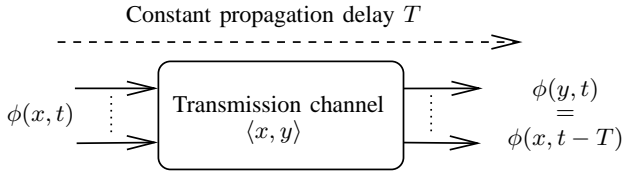}
  \caption{Transmission channel block}\label{fig:channel}
\end{figure}

The first element-model is derived in this section, namely the
model for lossless transmission channels with constant delay. They
exactly behave as transmission lines and a delay-based formulation
is provided. The derivation is rather straightforward but it is a
good example of application of Proposition~\ref{prop:ax1}.

\begin{result}
Given a lossless transmission channel corresponding to edge $E$ and having constant propagation delay $T>0$, the output flow is given by
\begin{equation}
  \phi(\eps(E),t)=\phi(\beta(E),t-T).
\end{equation}
The corresponding module is depicted in Fig. \ref{fig:channel}.
\end{result}
\begin{proof}
According to the conservation law (\ref{eq:cl}), the number of packets in transit $P_E(t)$ in the edge $E$ at time $t\in\mathbb{T}^u$ obeys
\begin{equation}
    P_E(t)=\int_{t_0(t)}^t\phi(\beta(E),s)ds
\end{equation}
where $t_0(t)=t-T$ since the propagation delay is constant. A packet sent at time $t-T$ will indeed be, at time $t$, still in the edge and about to leave. The result follows then from Proposition \ref{prop:ax1}.
%
\end{proof}

%
We are now in a position to introduce transmission channel
operators defining part of the network topology.
\begin{define}
The flow vectors $\phi(u^-,t)=\col_i[\phi(u_i^-,t)]$, $\phi(b^-,t)=\col_i[\phi(b_i^-,t)]$, $\phi(u^+,t)=\col_i[\phi(u_i^+,t)]$ and $\phi(b^+,t)=\col_i[\phi(b_i^+,t)]$ are related by transmission channel operators $\mathcal{R}_\bullet$, $\bullet\in\{ub,bu,bb\}$ as
\begin{equation}\label{eq:R}
  \begin{bmatrix}
    \phi(u^-,t)\\
    \phi(b^-,t)
  \end{bmatrix}=\mathcal{R}\begin{bmatrix}
     \phi(u^+,t)\\
     \phi(b^+,t)
  \end{bmatrix}+\mathcal{D}\delta(t)
\end{equation}
where
\begin{equation*}
\begin{array}{lclclcl}
  \mathcal{R}&=&\begin{bmatrix}
    0 & \mathcal{R}_{ub}\\
    \mathcal{R}_{bu} & \mathcal{R}_{bb}
  \end{bmatrix}&\mathrm{and}&
  \mathcal{D}&=&\begin{bmatrix}
    0\\
    \mathcal{D}_b
  \end{bmatrix}.
\end{array}
\end{equation*}
The matrices $\mathcal{R}_\bullet$, $\bullet\in\{ub,bu,bb\}$ correspond to routing matrices between output and input nodes. For instance, $\mathcal{R}_{ub}$ maps flows at user output nodes to flows at buffer input nodes. These matrices essentially consist of constant delay operators with delays corresponding to transmission channels. The full-rank input matrix $\mathcal{D}_b$ drives the vector of cross-traffic flows $\delta(t)$ to buffer input nodes.
\end{define}

\section{FIFO Buffer Model}\label{sec:buffer}

This section is devoted to the very important
buffer element which temporarily stores incoming information
before processing it. First, the standard fluid model for queues is
recalled \cite{Misra:00,Low:02,Srikant:04} and a complete delay-map is characterized. In order assign each input flow to its corresponding output flow and solve the output flow separation problem \cite{Ohta:98, Hespanha:01b, Farnqvist:02, Liu:04, Briat:10, Zhang:10}, the conservation law is then applied on the standard queue model, in a similar way as for transmission channels. The output
separation problem allows us to focus on the accurate description
of queues, which captures both the FIFO behavior and the form of
output flows. Some extra discussions and interpretations of the
results are also provided. Finally, a comparison to an existing
model for output flows is carried out and concludes in favor of
the proposed one.

\subsection{Queue model}

Routers have queues to store incoming packets temporarily. The
following integrator model \cite{Misra:00} can be proved to be a limit model of a
M/M/1 queue when the packet size and thus the processing time
tend to 0 \cite{Srikant:04, Moller:08}.
\begin{define}\label{ax:2}
\emph{The queue dynamics of buffer $b_i$ is governed by the model
\begin{equation}\label{eq:buffer}
  \dot{q}_i(t)=\sum_{j=1}^{\sigma(b_i)}\phi_j(b_i^-,t)-r_i(t)
\end{equation}
where the aggregate output rate is defined as
\begin{equation}\label{eq:outrate}
    r_i(t)=\left\{\begin{array}{lcl}
      c_i & &\mathrm{if}\ \mathcal{C}_i(t)\\
      \sum_j\phi_j(b_i^-,t) & & \mathrm{otherwise}.
    \end{array}\right.
\end{equation}
Above, $q_i$, $c_i$ and $\phi_j(b_i^-,t)$ represent the queue size, the maximal output capacity and the flow of type $j$ at the input respectively. The condition $\mathcal{C}_i(t)$ is given by
\begin{equation}
  \mathcal{C}_i(t):=\left(\left[q_i(t)>0\right]\vee\left[\sum_j\phi_j(b_i^-,t)>c_i\right]\right).
\end{equation}
The corresponding queuing delay can be easily deduced using the relation $\tau_i(t)=q_i(t)/c_i$.}
\mendprinciple
\end{define}

The above model can also be refined to capture additional features
such as finite maximal queue length, flow priorities, and multiple
output capacities. These extensions are omitted here since they
are straightforward. It is important to stress that this model is
incomplete and useless in this form. First, the output flow is
given in aggregate form. This prevents the modeling of the
appropriate routing of each output flow. Second, it does not
capture the queue FIFO behavior. Finally, the model does not
assign a specific queueing time to each input flow. In the
following subsection, the conservation law (\ref{eq:cl}) is used
in order to confer the FIFO property to the model and separate the
aggregate output flow into distinct flows.

\subsection{Forward and Backward Queuing Delays}\label{sec:delayop}

The maps defined in this section are very useful for obtaining a closed formula for the buffer output flows in Section \ref{sec:outputflow} and for RTT in Section \ref{sec:delayRTT}.

Let us consider first the buffer model (\ref{eq:buffer}) with
queueing delay $\tau_i(t)$. Assume that the time instants at which
packets enter the queue are chosen as reference times. We may then
be interested in predicting the packet output time. This leads to
the following definition:
\begin{define}[Forward delay operator]
  The forward delay operator $f_i:\mathbb{T}^u\to\mathbb{T}^u$ corresponding to buffer $b_i$ mapping, at a flow level, any input-time $t$ to the output-time $f_i(t)$ is defined as
\begin{equation}
  f_i(t):=t+\tau_i(t)
\end{equation}
where $\tau_i(t)$ is the queuing delay of buffer $b_i$.
\end{define}
It is easy to see that output-time can be readily computed from
the knowledge of input-time and queuing delay value. If however,
we would like to set the reference time to be the output time, we
may ask the question whether it is possible or not to retrieve the
input-time from it. This is equivalent to asking the question of
invertibility of the map $f_i$.
%
%

\begin{result}[\cite{Briat:10}]\label{res:gi}
  The map $f_i$ is invertible if and only if the input flow of the corresponding buffer is positive almost everywhere.\menlem
\end{result}
Hence, provided that there is a nonzero input flow to the buffer,
the input-time corresponding to a given output-time is
well-defined and can be obtained using the \emph{backward delay
operator}:
\begin{define}
   The backward delay operator $g_i:\mathbb{T}^u\to\mathbb{T}^u$ corresponding to buffer $b_i$ mapping, at a flow level, any output-time $t$ to the input-time $g_i(t)$ is defined as $g_i:=f_i^{-1}$ under the assumption of Result \ref{res:gi}.
\end{define}
We also have the following useful results:
\begin{result}[\cite{Briat:10}]\label{lem:fg}
The functions $f_i$ and $g_i:=f_i^{-1}$ obey:
    \begin{subequations}\label{eq:fg}
   \begin{eqnarray*}
    g_i^\prime(t)&=&\left\{\begin{array}{ll}
    \displaystyle{c_i\left(\sum_{k=1}^{\eta(b_i)}\phi_k(b_i^-,g_i(t))\right)^{-1}} &\mathrm{if\ }\mathcal{C}_i(g_i(t))\\
    1 &\mathrm{otherwise}
  \end{array}\right.\label{eq:prop3}\\
    f_i^\prime(t)&=&\left\{\begin{array}{lcl}
    \displaystyle{c_i^{-1}\sum_{k=1}^{\eta(b_i)}\phi_k(b_i^-,t)} &&\mathrm{if\ }\mathcal{C}_i(g_i(t))\\
    1 &&\mathrm{otherwise}
  \end{array}\right.\label{eq:prop4}\\
  {[\tau_i(g_i(t))]}^\prime&=&\left\{\begin{array}{lcl}
    1-\dfrac{c_i}{\sum_{k=1}^{\eta(b_i)}\phi_k(b_i^-,g_i(t))}&\mathrm{if\ }\mathcal{C}_i(g_i(t))\\
    1&\mathrm{otherwise}
  \end{array}\right.\label{eq:prop5}
  \end{eqnarray*}
\end{subequations}
where $f^\prime(t)$ stands for the the upper right Dini derivative of $f(t)$, i.e. ${f^\prime(t)=\limsup_{h\downarrow0}h^{-1}\left(f(t+h)-f(t)\right).}$\menlem
\end{result}

The following technical result allows to simplify the conditions involved in hybrid models:
\begin{result}\label{res:equiv}
  The equivalence $\mathcal{C}(g_i(t))\Leftrightarrow\mathcal{C}(t)$ holds.
\end{result}
\begin{proof}
\textbf{Proof of $\Rightarrow$:} If the buffer is congested at time $g_i(t)$ then the buffer will also be congested at time $t$ since the data entered at time $g_i(t)$ leave at time $t$.\\
\textbf{Proof of $\Leftarrow$:} Conversely, if there is any data to leave at time $t$, they  must have entered in the queue in the past, i.e. at time $g_i(t)$. Equivalence is proved.
\end{proof}

\subsection{FIFO Buffer Output Flow Separation}\label{sec:outputflow}

In this section, we use the results given above and the
conservation law in order to improve the buffer
modeling by adding the FIFO characteristics and splitting the
aggregate output flows into a sum of distinct ones. Without
further consideration on the queue type, there exists an infinite
number of ways to separate the aggregate output flow directly from
the queuing model of Definition \ref{ax:2}. When a FIFO queue
(i.e. order preserving) is considered, it turns out that the
output flow separation problem is easily solvable. The FIFO
characterization and output flow separation problems have been
fully solved in \cite{Briat:10}. In this section, we will simply
recall and explain these results, and connect them to the
conservation law (\ref{eq:cl}).
\begin{result}[\cite{Briat:10}]
  Let us consider the queueing model (\ref{eq:buffer}) which we assume to represent a FIFO queue. The output flow corresponding to the input flow $\phi_\ell(b_i^-,t)$, $\ell\le\eta(b_i^-)$ is given by
  \begin{equation}\label{eq:bufferoutput}
  \begin{array}{lcl}
    \phi_\ell(b_i^+,t)&=&g_i^\prime(t)\phi_\ell(b_i^-,g_i(t))\\
    &=&\left\{\begin{array}{lcl}
      \dfrac{c_i\phi_\ell(b_i^-,g_i(t))}{\sum_j^{\eta(b_i)}\phi_j(b_i^-,g_i(t))}&&\mathrm{if\ }\mathcal{C}_i(t)\\
      \phi_\ell(b_i^-,t)&&\mathrm{otherwise}
    \end{array}\right.
  \end{array}
  \end{equation}
\end{result}
\begin{proof}
  A proof is available in \cite{Briat:10}. A more direct one based on Proposition \ref{prop:ax1} is given here. Noting that for buffer $b_i$, we have $t_0(t)=g_i(t)$, then using Proposition (\ref{prop:ax1}) and the formulas of Result \ref{lem:fg}, we get equation (\ref{eq:bufferoutput}) with the difference that the condition is $\mathcal{C}(g_i(t))$. However, from Result \ref{res:equiv}, the condition $\mathcal{C}(g_i(t))$ is equivalent to $\mathcal{C}(t)$ and the result follows.
\end{proof}

The same model has been also proposed in \cite{Ohta:98, Liu:04} but claimed without any proof. We have shown above that this model is an immediate consequence of the information conservation law and gives, for the first time, a theoretical proof for it. This considerably strengthens the trust we may have in this model. A comparison with packet-level simulations in the next subsection tends to show its exactness.

This model also deserves interpretation: formula (\ref{eq:bufferoutput}) says that output flows consist of scaling and shifting of the input flows. The delay accounts for high flow viscosity and captures the queue FIFO behavior, \emph{at a flow level}, while the nonlinear ratio expresses the flow coupling at the core of the flow and clock-coupling phenomena, see section \ref{sec:fluidflow}, since each output flow depends on the corresponding input flow and all the other ones as well. A change in a single flow will affect all the output flows. This model also tells that the output flow corresponding to the input flow $\phi_\ell(b^-,t)$ is expressed as a (delayed) ratio of the input flow $\phi_\ell(b^-,t)$ to the total input flow that entered the buffer at the same time. Hence, the output flows are proportional to relative flows modeling the 'chance' of having a packet of certain type served at time $t$. This 'chance' is then scaled-up by the maximal output capacity to utilize the available capacity. 

Operators representing buffers can now be introduced:
\begin{define}
  The buffer operator $\mathcal{B}_i$ with $\eta(b_i)$ input flows is defined as
\begin{equation}
  \begin{array}{lcrclc}
  \mathcal{B}_i&:&
  \phi(b_i^-,t)&\to&\phi(b_i^+,t)
  \end{array}
\end{equation}
where the output flows and the buffer state are governed by (\ref{eq:buffer}) and (\ref{eq:bufferoutput}). Using these operators, we can build a matrix of operators $\mathcal{B}$ connecting the $\phi(b^-,t)$'s to the $\phi(b^+,t)$'s as
\begin{equation}
    \phi(b^+,t)=\mathcal{B}\phi(b^-,t)
\end{equation}
where $\mathcal{B}=\diag_i\left\{\mathcal{B}_i\right\}$.
\end{define}

\begin{figure}
  \centering
        \psfrag{d}[c][c]     {\small{Forward queueing delay $\tau$}}
        \psfrag{b}[c][c]     {\small{Backward queueing delay $\tau(g)$}}
        \psfrag{ch}[c][c]     {\small{Queue $\langle b^-,b^+\rangle$}}
        \psfrag{ip}[c][c]     {\small{$\phi(b^-,t)$}}
        \psfrag{op}[c][c]     {\small{\shortstack{$\phi(b^+,t)$\\defined in (\ref{eq:bufferoutput})}}}
  \includegraphics[width=0.4\textwidth]{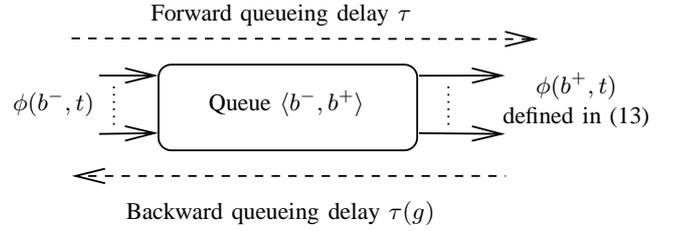}
  \caption{Queue/Buffer block}\label{fig:buffer}
\end{figure}

\subsection{Comparison with Another Model}\label{sec:compbuffer}

Two main models for buffer output flows have been reported in the literature on fluid-flow models: the flow-based model \cite{Ohta:98, Liu:04, Briat:10} described in this paper and the pseudo-queue-based one \cite{Hespanha:01b,Farnqvist:02,Zhang:10} given by
\begin{equation}\label{eq:modelq}
\phi_\ell(b_i^+,t)=\left\{\begin{array}{lcl}
    \dfrac{q_i^\ell(t)c_i}{\sum_kq_i^k(t)}&&\mathrm{if}\ q_i(t)\ne 0\\
    0 && \mathrm{otherwise}
\end{array}\right.
\end{equation}
where $q_i^\ell(t)$ is the number of packets of type $\ell$ in
queue $i$ and $c_i$ is the maximal output capacity of queue $i$.

Until now, these models have not been confronted to each others. In the following, they will be theoretically and experimentally compared, and it will be shown that the flow-based model is the only model that faithfully characterizes the actual output flows, validating then the proposed conservation-law-based paradigm.

\subsubsection{Theoretical argumentation}

First, the above model assumes that the 'chance' of having a packet of type $\ell$ at the output at time $t$ is $q_i^\ell(t)/\sum_iq_i^\ell(t)$. Model (\ref{eq:modelq}) then makes no difference in picking a packet in the middle, at the end or at the beginning of the queue since only the number of packets matters. It is thus unable to capture the FIFO characteristic of the queue since swapping packets in the queue does not modify the output flow. In contrast, the proposed model does capture this characteristic through the delay dynamical model and the delayed nonlinear input-output relationship involving flows directly: relative variations of the input flows are passed to the output flows after some queueing delay. Note however that both models coincide at equilibrium.

Second, since output flows in model (\ref{eq:modelq}) are computed
from the integration of input flows, it turns out that the map
from input flows to output flows is a nonlinear low-pass filter
with 'bandwidth' equal to $1/q(t)$ when $q(t)>0$. High frequencies
in the input flows are hence filtered out, making the existing
model inaccurate for high frequency flows (fast transient),
especially when the queue size is large. Note that the actual buffer behavior does not have any filtering effect, it just behaves as an ordered tank. On the other hand, the
proposed model does not filter out any frequency band due to its
feedthrough structure. It however has a distortion effect on the
input flows due to the nonlinear structure and the dynamically
changing delay, i.e. change of frequency and amplitude. As a result, the queue-based model does not satisfy any conservation law since low-pass filters dissipate energy all over the frequency band, resulting in information loss at the model level. This is in total contradiction with the actual queue behavior that just stores information and does not dissipate anything. Note however that the flow-based model intrinsically satisfies the conservation since the model is derived from it.

Last, the proposed model incorporates naturally the queuing delay in the expression, while for model (\ref{eq:modelq}) it is unclear which delay to consider, since the order of arrival of data is not tracked.

\subsubsection{Case Analysis}

To compare the models, let us consider two input flows given by
\begin{equation}
  \begin{array}{lcl}
    \phi_1(t)&=&(1+\beta)c\left(1+\Sq(\omega t)\right)/2,\\
    \phi_2(t)&=&(1+\beta)c\left(1-\Sq(\omega t)\right)/2
  \end{array}
\end{equation}
where $c>0$ and $\omega>0$ are the buffer output capacity and the
oscillation of flows. The term $\beta>0$ is a tunable parameter
related to the amplitude of the inputs flows and the function
$\Sq(\omega t):=\sign(\sin(\omega t))$ is a square function of
period $T:=2\pi/\omega$. Since the flows are in phase opposition, they lead to an alternation of packet types in the queue while packet populations remain roughly close to each other at any time. Therefore, the output flows should reflect the actual content of the queues and the model should be able to keep track of the order of arrival of packets in the queue. 

The queue-based model (\ref{eq:modelq}) predicts the queues
\begin{equation}
  q_i(t)=\theta(t)q_i(0)+\int_0^t\theta(t-s)\phi_i(s)ds,\ i=1,2
\end{equation}
with $\theta(t) = \left(\dfrac{q(0)}{q(0)+\beta ct}\right)^{1/\beta}$
from which it is quite difficult to foresee the shape of the
output flows. We can however note that the filtering effect of the convolution operator with kernel $\theta$ will deform the input flows, making then the output flows not square anymore.

When the proposed flow-based model is considered, it is enough
to compute the forward and backward delays, and apply the formula
for output flows:
\begin{equation}
  \begin{array}{lcl}
    \tau(t)&=&\beta t+\tau(0),\\
    f(t)&=&(1+\beta)t+\tau(0),\\
    g(t)&=&\dfrac{1}{1+\beta}(t-\tau(0)),\\
    \phi_1(b^+,t)&=&c(1+\Sq(\omega g(t)))/2,\\
             &=&\dfrac{c}{2}\left[1+\Sq\left(\dfrac{\omega}{1+\beta}(t-\tau(0))\right)\right],\\
    \phi_2(b^+,t)&=&\dfrac{c}{2}\left[1-\Sq\left(\dfrac{\omega}{1+\beta}(t-\tau(0))\right)\right].
  \end{array}
\end{equation}
In this case, the predicted output flows have the same shape as the input flows but with a different frequency and amplitude. The proposed output flow model then exactly captures the content of the queue, that is the alternation of blocks of size $N(T)$ of type 1 and type 2. It is also immediate to see that the number of packets $N(T)=N_i(T),\ i=1,2$ received by the buffer over one period $T$ is given by 
\begin{equation}
  N(T):=\int_0^T\phi_1(b^+,s)ds=\dfrac{\pi(\beta+1)c}{\omega}.
\end{equation}

By virtue of the information conservation law, the same number of packets is retrieved at the output over the period $(\beta+1)T$, enlarged due to the limiting output capacity $c$. It is quite convincing that the flow-based model yields a much more coherent picture for this example.

\subsubsection{Simulation}

\begin{figure}[H]
\begin{minipage}[b]{0.49\linewidth}
\centering
 \includegraphics[width=\textwidth]{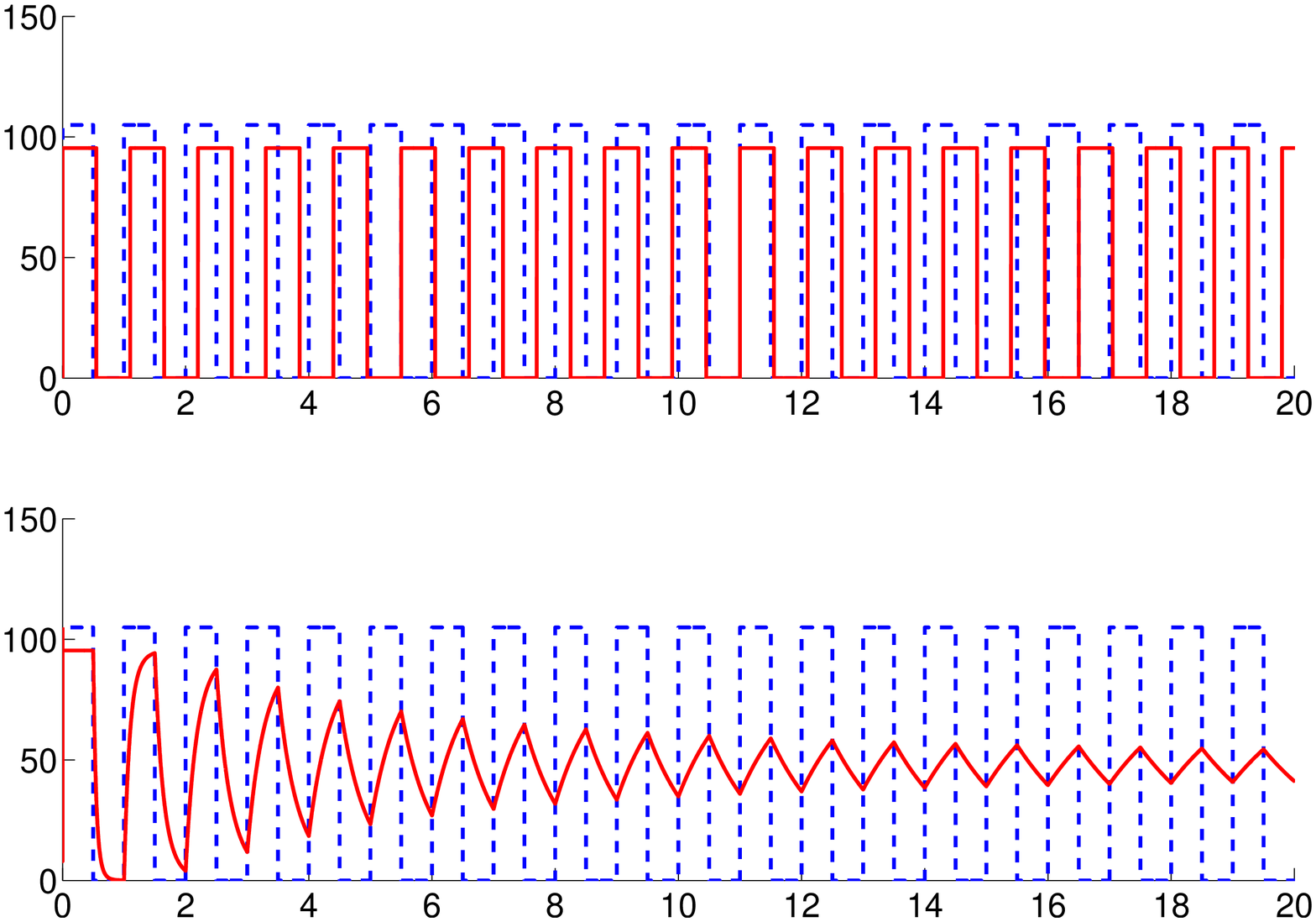}
\caption{Comparison of the output flows predicted by model (\ref{eq:bufferoutput}) (top) and model (\ref{eq:modelq}) (bottom). Plain: output flow $\phi_1(b^+,t)$, dashed: input flow $\phi_1(b^-,t)$}\label{fig:compflow}
\end{minipage}
\hfill
\begin{minipage}[b]{0.49\linewidth}
\centering
  \includegraphics[width=\textwidth]{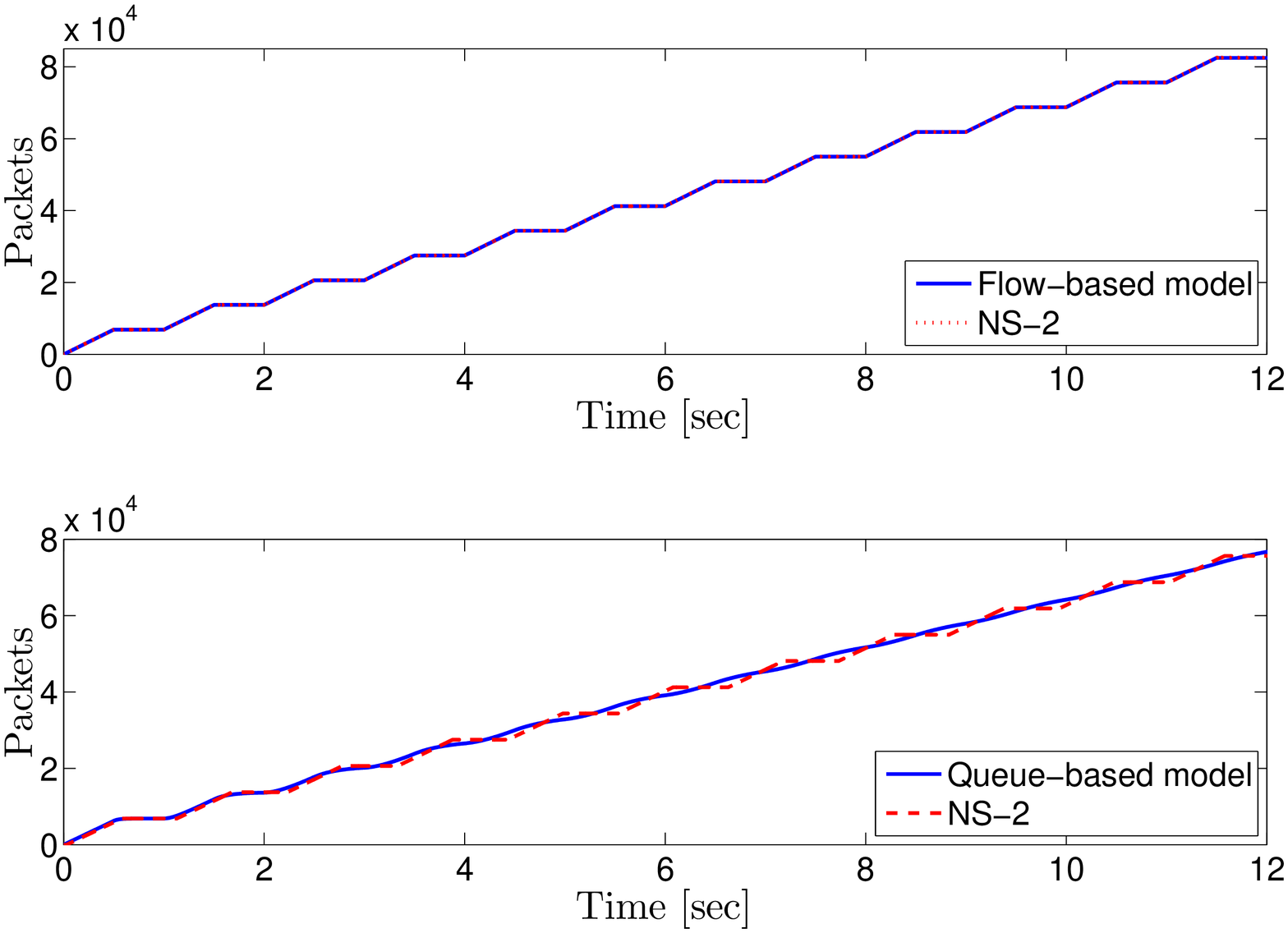}
\caption{Comparison with NS-2 simulations of the total number of packets counted at the inputs of the buffers described by the proposed flow based model (\ref{eq:buffer})-(\ref{eq:bufferoutput}) and the queue based model (\ref{eq:buffer})-(\ref{eq:modelq}).}\label{fig:flowNS2}
\end{minipage}
\end{figure}

For simulation, we choose $\omega=2\pi$ (i.e. $T=1$s), $\beta=1$, $c=100$Mb/s and $\tau(0)=0$. The output flows obtained from the different models are depicted in Fig.~\ref{fig:compflow}, where we observe the behaviors predicted by the above calculations. Notably, the output flows predicted by model (\ref{eq:modelq}) tend to slow-down (low pass-filtering effect), decrease along time and seem to both tend to $c/2$, which is basically unrepresentative of the actual content and output of the queue. We can also notice the decrease of the bandwidth for the queue-based model as long as the queue grows in size.

For comparison with NS-2, which deals with packets rather than flows, we integrate (up to an additional constant) the output flows predicted by each model to obtain a number of packets so that the comparison with NS-2 makes sense. The results are depicted in Fig. \ref{fig:flowNS2} where we can see that the proposed model yields exactly the same results as NS-2 while the queue-based model is unable to track the stair-like curve returned by NS-2. It is also important to stress that the considered scenario is quite convenient for the queue-based model since the input flows contain mostly constant parts (low frequency parts). A very fluctuating input flow would be very penalizing and would make the low-pass filtering effect of the queue-based model even more apparent.

\section{Complete user model - Window control}\label{sec:user}

The derivation of the user model is, partially, still an open question and a
complete solution, based on the conservation law (\ref{eq:cl}), is
proposed in this section. We assume that the implemented
congestion control protocol admits a fluid-flow approximation, for
instance interpolating the discrete-time trajectories of the real
protocol \cite{Misra:00,Jacobsson:09}. The conservation law is
then applied over the circuit used by a user to obtain the
so-called ACK-clocking model \cite{Jacobsson:08b,Tang:10} relating
flight-size, user sending rate and RTT together. Several
expressions for RTT are provided according to the considered
reference time, similarly as for buffers. These results are
finally merged together in order to clarify the connection between
the flight-size and the user sending rate. The last step concerns
the derivation of formulas relating the above variables to user
congestion window size.

\subsection{Protocol model}

Here we assume that the congestion control algorithm can be
represented as a continuous-time (hybrid) dynamical system. That
is we have the following fact:
\begin{fact}
  There exist bounded functionals $\mathcal{P}_i$, $\mathcal{W}_i$ and $\mathcal{U}_i$ such that the trajectories $(z_i(t),w_i(t))$ of the following continuous-time model defined over $\mathbb{T}^u$
\begin{equation}\label{eq:protocol}
\begin{array}{lcl}
    \dot{z}_i(t)&=&\mathcal{P}_i(z_i(t),\mu_i(t))\\
    w_i(t)&=&\mathcal{W}_i(z_i(t),\mu_i(t))\\
    \phi_i(u_i^+,t)&=&\mathcal{U}_i(w_i(t),\phi_i(u_i^-,t))
\end{array}
\end{equation}
match the trajectories of the asynchronous protocol (defined on
$\mathbb{T}_i$) at points in $\mathbb{T}^u\cap\mathbb{T}_i$.
Above, $z_i$, $\mu_i$, $\phi_i(u_i^-,\cdot)$ and
$\phi_i(u_i^+,\cdot)$ are the state of the protocol, the
congestion measure, the acknowledgment flow rate and the user
sending flow respectively. The window size, $w_i$, is considered
as the number of outstanding packets to track and supposed to be
(weakly) differentiable.
\end{fact}

A procedure to solve the above interpolation problem has been first proposed in \cite{Misra:00} for TCP and has been adapted to FAST-TCP in \cite[Appendix C.]{Jacobsson:09}.

\subsection{The ACK-Clocking model}\label{sec:ack}

The ACK-clocking model \cite{Jacobsson:08b,Tang:10} is a very
important consequence of the conservation-law. It characterizes the \emph{flight-size}\footnote{The number of
outstanding packets.} $\digamma_i(C_i,t):=P_{C_i}(t)$ of the user
$u_i$ at any time $t\in\mathbb{T}^u$ over a circuit $C_i=\langle
u_i^+,u_i^-\rangle$. The importance of the ACK-clocking model lies
in the semantic; it adds to the model by relating RTT, flow and
flight-size together\footnote{Note however that in \cite{Tang:10}
the window size is considered instead of the flight-size, which is
rather different. To make the distinction, the window-based ACK clocking model is denoted by W-ACK while the fligh-size-based one by FS-ACK. Equivalence holds when some conditions, such as
'flight-size$\ge$window size', are met. We will come back on this
in Section \ref{sec:userwinflight}.}. With this result, it is
possible to incorporate a number of important user properties and
mechanisms into the corresponding model.

\begin{result}[FS-ACK-Clocking]
The (FS)ACK clocking model is given by
\begin{equation}
\begin{array}{lcl}
      \digamma_i(C_i,t+\RTT_i\{t\}) &=& \displaystyle{\int_{C_i}\phi(\theta,t+\RTT_i\{t\})d\theta}\\
                        &=& \smashoperator{\int_{t}^{t+\RTT_i\{t\}}}\phi_i(u_i^+,s)ds
\end{array}
\end{equation}
where $\RTT_i\{t\}$ is the RTT of a packet sent at time $t$ by user $u_i$ over the circuit $C_i$.\menlem
\end{result}
\begin{proof}
  Since flight-size is a number of packets in a circuit, it can be cast as a spatial integration over the corresponding circuit. Thus, according to the conservation-law, it is possible to convert the spatial integration into a temporal one provided that we can determine the integration bounds. To obtain them, we use the notion of RTT and suppose that a data sent by user $u_i$ in the circuit $C_i$ at time $t$ has a round-trip-time given by $\RTT_i\{t\}$. This means that the packets sent between $t$ and $t + \RTT_i\{t\}$ are unacknowledged at $t+\RTT_i\{t\}$ and thus still in the circuit. Hence, the corresponding temporal integral has bounds $t$ and $t+\RTT_i\{t\}$.
\end{proof}

\subsection{Round-Trip-Time Models}\label{sec:delayRTT}

The RTT consists of sum of a constant and a time-varying part,
namely the propagation delays and the queuing delays, which have
been characterized in Sections \ref{sec:transmed} and
\ref{sec:buffer}, respectively. By combining these results
together, it is immediate to obtain a RTT based forward model on a
\emph{forward circuit operator}, which considers the packet input
time as a reference.

To properly define it, let us consider a circuit $C$ with $N$ queues, indexed from 1 to $N$. The indices $0$ and $N+1$ are used to denote the input and output nodes of the circuit respectively. Given a packet input time $t$, the corresponding packet output time $t_C$ and RTT obey the following formulas based on the forward circuit operator $\mathscr{F}_C$.
\begin{define}
  The forward circuit operator $\mathscr{F}_C:\mathbb{T}^u\to\mathbb{T}^u$ of circuit $C$ assigned to any packet at input time $t$, an output-time $t_C(t)$ is given by
\begin{equation}\label{eq:FRTT}
\begin{array}{lcl}
    t_C(t)&=&\mathscr{F}_C(t),\\
    \mathscr{F}_C&=& \nabla^{-1}_{N,N+1}\circ f_{N}\circ\nabla^{-1}_{N-1,N}\circ f_{N-1}\circ \ldots\\
    &&\circ\nabla^{-1}_{2,3}\circ f_{2}\circ\nabla^{-1}_{1,2}\circ f_{1}\circ\nabla^{-1}_{0,1}
\end{array}
\end{equation}
where $\nabla_{i,j}$ is the constant delay operator with delay corresponding to the propagation delay between $i$ and $j$ and $\circ$ the composition operator. The corresponding RTT expression is then given by
\begin{equation}
\mathrm{RTT}\{t\}=t_C(t)-t=(\mathscr{F}_C-\mathrm{Id})(t)
\end{equation}
where $\mathrm{Id}$ is the identity operator.
\end{define}
Formula (\ref{eq:FRTT}) actually represents the alternation between constant delay operators corresponding to transmission channels delay (the $\nabla_i$'s) and the queuing delays corresponding to queues (the $f_i$'s). Example \ref{ex:RTT} illustrates this formula on the topology depicted in Fig. \ref{fig:graph}. The same formulas, albeit expressed in different ways, have been also obtained in \cite[Section 3.3.5]{Jacobsson:08} and \cite[equations (7d-7f)]{Tang:10}.

Although immediate to obtain, these expressions suffer from
several drawbacks. First, operator $\mathscr{F}_C$ is clearly
noncausal since it requires the knowledge of future information.
Second, as pointed out in Section \ref{sec:delayop}, the most
convenient reference-time to use is the output time. In the user
modeling problem, it coincides with the reception time of
acknowledgments and the moment when the user receives the RTT
information. Hence, it seems to be more convenient to consider a
\emph{backward circuit operator} based on the backward delay
operator of Section \ref{sec:delayop}. The existence of such an
operator is immediately inferred from the existence of the
backward delay operator.
\begin{define}
   The backward circuit operator $\mathscr{B}_C:\mathbb{T}^u\mathbb{T}^u$ of circuit $C$ assigned to any packet at output time $t_C$, an input-time $t(t_c)$ is given by
\begin{equation}
\begin{array}{cll}
    \mathscr{B}_C&:=&\mathscr{F}_C^{-1},\\
    t(t_C)&=&\mathscr{B}_C(t_C),\\
    \mathscr{B}_C&=&\nabla_{0,1}\circ g_{1}\circ\nabla_{1,2}\circ g_{2}\circ\nabla_{2,3}\circ\ldots\\
    &&\circ g_{N-1}\circ\nabla_{N-1,N}\circ g_{N}\circ\nabla_{N,N+1}.
\end{array}
\end{equation}
Moreover, the corresponding RTT expression is given by
\begin{equation}
      \mathrm{RTT}\{t\}=t_C-t(t_C)=(\mathrm{Id}-\mathscr{B}_C)(t_C).
\end{equation}
\end{define}

It is important to stress that the RTT formula given above looks noncausal since the RTT of a packet sent at time $t$ is defined in
terms of time $t_C>t$. This is however not a problem since the
RTT information is only available, and then used, by the user at
time $t_C$, when the ACK packet is actually received. What is
important is the causality of the operator $\mathscr{B}_C$ in
order to ensure computability of the RTT at any time. This emphasizes
once again the relevance of considering output times as
references. The example below illustrates the above discussions:
\begin{example}\label{ex:RTT}
  In the single-user/single-buffer case, the above expressions reduce to
  \begin{equation}
  \begin{array}{lcl}
        t_C(t)&=&t+T_f+T_b+\underbrace{\tau(t+T_f)}_{\mbox{Future information}}\\
        t(t_C)&=&\underbrace{g(t_C-T_b)}_{\mbox{Past information}}-T_f\\
        \RTT\{t\}&=&t_C-t\\
        &=&t_C-g(t_C-T_b)+T_f\\
        &=&T_b+T_f+\tau(g(t_C-T_b))
  \end{array}
  \end{equation}
  where $T_f$ and $T_b$ are the forward and backward propagation delays corresponding to the constant delay operators $\nabla_{0,1}$ and $\nabla_{1,2}$.
\end{example}

Using the backward expression of RTT and the relation
$\mathscr{F}_{C_i}\circ\mathscr{B}_{C_i}=\mathscr{B}_{C_i}\circ\mathscr{F}_{C_i}=\mathrm{Id}$,
it easy to obtain the following result:
\begin{result}\label{res:ACK_clocking_model}
  The flight size obeys
  \begin{eqnarray}
      \digamma_i(C_i,\mathscr{F}_{C_i}(t)) &=& \smashoperator{\int_{t}^{\mathscr{F}_{C_i}(t)}}\phi(u_i^+,s)ds\label{eq:FS1}\\
      \digamma_i(C_i,t) &=& \smashoperator{\int_{\mathscr{B}_{C_i}(t)}^{t}}\phi(u_i^+,s)ds.\label{eq:FS2}
  \end{eqnarray}
  \mendprop
\end{result}
\begin{proof}
  This is an immediate consequence of the conservation-law (\ref{eq:cl}) (through the ACK-clocking model) and the above RTT models.
\end{proof}
%
%
%
%
%
Note that in \cite{Tang:10}, a model is obtained directly from the
W-ACK-clocking applied directly to a selected topology. However,
the model is not modular itself since hand calculations are needed
to make it implementable that results unfortunately in complexity
very sensitive to the topology. Hence the objective of obtaining a
metamodel is not attained. The reason is that the ACK-clocking
model is used in \emph{implicit form} while our proposed method
incorporates solutions of it, yielding an \emph{explicit
formulation} achieving the characteristics of a metamodel, i.e.
modularity and scalability.

\subsection{ACK-Clocking Dynamics and User Flow Computation}\label{sec:ackflow}

Since the flight-size expression (\ref{eq:FS2}) is exactly of the form (\ref{eq:cl}), then Proposition \ref{prop:ax1} can be immediately applied to derive an explicit expression for the output flow of a given circuit, that is the flow of received acknowledgments:
\begin{result}
Let us consider a circuit $C_i=\langle u_i^+,u_i^-\rangle$. Then,
the ACK-flow that user $u_i$ receives is given by
  \begin{equation}\label{eq:ackflow}
    \phi(u_i^-,t)=\mathscr{B}^\prime_{C_i}(t)\phi_i(u_i^+,\mathscr{B}_{C_i}(t)).
  \end{equation}
  \mendprop
\end{result}
\begin{proof}
The key idea is to remark that
$\digamma_i(C_i,t)=N_{u_i^+}(t,\mathscr{B}_{C_i}(t))$. Hence,
using Proposition \ref{prop:ax1} and noting that the ACK-flow
corresponds to the flow leaving the circuit $\phi(u_i^-,t)$, the
result is obtained. Differentiability of $\mathscr{B}_{C_i}(t)$ is inferred from the differentiability of the backward delay operators.
%
%
\end{proof}
Note that differentiation of (\ref{eq:FS2}) yields
\begin{equation}\label{eq:mdrlol1}
  \phi_i(u_i^+,t)=\digamma_i^\prime(C_i,t)+\mathscr{B}^\prime_{C_i}(t)\phi_i(u_i^+,\mathscr{B}_{C_i}(t))
\end{equation}
meaning that to maintain a constant flight size, i.e.
$\digamma_i^\prime(C_i,t)=0$, the user has to naturally send data
at the same rate it receives ACK packets: \emph{this is exactly
ACK-clocking but expressed at a flow level}. By flow integration,
we can easily recover the 'packet-level ACK-clocking'. A similar
expression is reported in \cite{Zhang:10}, but stated without proof and using the model in (\ref{eq:modelq}) to represent the buffer output flows. Once again the proposed methodology allows to provide strong mathematical foundations for some existing results.

\subsection{User Flow, Flight-Size and Congestion Window Size}\label{sec:userwinflight}

We need to clarify the relation between a user congestion window
size $w_i(t)$ and its sending rate $\phi(u_i^+,t)$. First, recall
that the congestion window size corresponds to the desired
flight-size, while the flight-size is the current number of
packets in transit. The window size is then a \emph{reference} to
track while the flight size is the \emph{controlled output}. The
\emph{control input} is the user sending rate.

When the window size increases, the user can immediately send a
burst of packets to equalize the flight- and window-sizes. In such
a case, we can ideally assimilate them to be equal\footnote{This is the main assumption in \cite{Tang:10} justifying the use of the W-ACK-clocking model.} (and so are
their derivatives). The small delay corresponding to the protocol
reaction time can be easily incorporated in the constant part of
the RTT. The problem is, however, slightly more difficult when the
congestion window size becomes smaller than the flight-size. In
such a case, we can not withdraw packets from the network and the
only thing we can do is to wait for the packets in the network to
be acknowledged until, at some point, the flight size becomes
equal to the window size. This basically means that the rate of
decrease of the flight-size is equal to the rate of received
acknowledgments (the rate at which data leave the network).
Therefore, while positive slope of the flight-size is ideally
unconstrained from above, the negative slope is lower bounded. In
\cite{Jacobsson:08}, a rate-limiter is a posteriori added to the model in order to constrain the
negative slope of the flight size. This solution is however difficult to implement in the context of \cite{Jacobsson:08} due to the time-varying nature of the slope lower bound and the
lack of any ACK-flow model. Note also that in most recent works
\cite{Zhang:10,Tang:10}, this problem is automatically excluded by
considering that the flight-size is always smaller than the window
size, and that the window size does not decrease `too-much'. To
the authors' best knowledge no well-rounded solution has been provided
yet for the problem of congestion window size
decrease. We provide below an explicit and complete
solution to this problem, regardless of the rate of the variation
of congestion window size. This is achieved through an
augmentation of the user model and the consideration of the flow of ACK packets.

\begin{figure}[H]
  \centering
        \psfrag{ch}[c][c]     {\small{User $\langle u^-,u^+\rangle$}}
        \psfrag{m}[c][c]     {\small{$\mu(t)$}}
        \psfrag{ip}[c][c]     {\small{\shortstack{$\phi(u^-,t)$\\ defined in (\ref{eq:ackflow})}}}
        \psfrag{op}[c][c]     {\small{\shortstack{$\phi(u^+,t)$\\ defined in (\ref{eq:userflow1})}}}
  \includegraphics[width=0.35\textwidth]{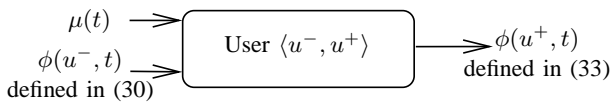}
  \caption{User block}\label{fig:user}
\end{figure}

According to the above discussion, the flight-size must obey
\begin{equation}\label{eq:flightsizecontrolled}
    \digamma_i^\prime(C_i,t)=\left\{\begin{array}{lcl}
      \dot{w}_i(t) && \mathrm{if\ }\mathcal{T}_i(t)\\
      -\phi(u_i^-,t)&& \mathrm{otherwise}
    \end{array}\right.
\end{equation}
where $\mathcal{T}_i(t)$ is a condition which is true when no
lower limit on the rate of variation of the flight-size is imposed
and false otherwise.

\begin{result}
The flight-size $\digamma_i(C_i,t)$ satisfies (\ref{eq:flightsizecontrolled}) if the user sending rate is defined as
  \begin{equation}\label{eq:userflow1}
    \phi(u_i^+,t)=\left\{\begin{array}{lcl}
      \dot{w}_i(t)+\phi(u_i^-,t) && \mathrm{if\ }\mathcal{T}_i(t)\\
      0&& \mathrm{otherwise}
    \end{array}\right.
  \end{equation}
  where $\mathcal{T}_i(t)=\left([\pi_i(t)=0]\wedge[\dot{w}_i(t)+\phi(u_i^-,t)\ge0]\right)$ and
  \begin{equation}\label{eq:userflow2}
    \dot{\pi}_i(t)=\left\{\begin{array}{lcl}
      0 && \mathrm{if\ }\mathcal{T}_i(t)\\
      \dot{w}_i(t)+\phi(u_i^-,t) && \mathrm{otherwise}.
    \end{array}\right.
  \end{equation}
  Moreover, this model is the simplest one.\menlem
\end{result}
\begin{proof}
  The ACK-buffer $\pi_i$, taking nonpositive values, measures the number of ACK packets to retain in order to balance the flight- and window-sizes. When the virtual buffer has negative state, i.e. $\pi_i(t)<0$, the arriving ACK-packets have to be retained until the state reaches 0. Once zero is reached, the user can start sending again until the window size decreases too fast, i.e. $\dot{w}_i(t)<-\phi(u_i^-,t)$. Substitution of the user sending rate defined by (\ref{eq:userflow1}) and (\ref{eq:userflow2}) in (\ref{eq:mdrlol1}) yields the flight-size behavior (\ref{eq:flightsizecontrolled}). To see that the model is minimal, it is enough to remark that both conditions in $\mathcal{T}_i(t)$ are necessary.
\end{proof}

In order to characterize the ACK-retaining mode, the ACK-buffer
(\ref{eq:userflow2}) has to be adjoined to the protocol model
(\ref{eq:protocol}), resulting in an augmentation of the state of
the user model. The protocol behavior depends on the measurements
$\mu_i(\varkappa_t)$ which are functions of the overall network
state $\varkappa_t$; this state is discussed in more detail in
Section \ref{sec:general}.

We are now in a position to define user operators from
(\ref{eq:userflow1}).
\begin{define}
  The user operator $\mathcal{U}_i(w_i):\mathbb{R}_+\to\mathbb{R}_+$ mapping the ACK-flow $\phi(u_i^-,t)$ to the sending flow $\phi(u_i^+,t)$ is given by
\begin{equation}
  \phi(u^+,t)=
   \mathcal{U}(w)
  \phi(u^-,t)
\end{equation}
where $\mathcal{U}(w)=\diag_i\left\{\mathcal{U}_i(w_i)\right\}$ and $\mathcal{U}_i$ is given in (\ref{eq:protocol}).
\end{define}

\section{General Network model}\label{sec:general}

Modular and independent models for transmission channels, buffers
and users have been developed in Sections \ref{sec:transmed},
\ref{sec:buffer} and \ref{sec:user} respectively. In this section
we summarize the obtained results in a compact form involving
dynamical systems and operators, and properties of the model are discussed. Notably, correspondence of the proposed model with existing ones is emphasized/recalled.

\subsection{General model}

The general network model takes the form
\begin{equation}\label{eq:generalmodel}
\begin{array}{rcl}
    \dot{\varkappa}(t)&=&\mathcal{N}\left(\varkappa_t,\phi(u^-,t),\phi(b^-,t)\right)\\
    \digamma_i(C_i,t)&=&\int_{\mathscr{B}_{C_i}(t)}^t\phi(u_i^+,s)ds
\end{array}
\end{equation}
with
%
\begin{equation}\label{eq:topology}
\Phi(t)=\begin{bmatrix}
  0 & 0 & \mathcal{U}(w) & 0\\
  0 & 0 & 0 & \mathcal{B}\\
  0 & \mathcal{R}_{ub} &  0 & 0\\
  \mathcal{R}_{bu} & \mathcal{R}_{bb} & 0 & 0
\end{bmatrix}\Phi(t)+\begin{bmatrix}
  0\\
  0\\
  \hline
  0\\
  \mathcal{D}_b
\end{bmatrix}\delta(t)
\end{equation}
where $\Phi(t)=\col(\phi(u^+,t),\phi(b^+,t),\phi(u^-,t),\phi(b^-,t))$ and $\varkappa=\col(\tau,z,\pi)$ are the flows and the state of the network, respectively. The hybrid models for user and queue dynamics are described by the
nonlinear (discontinuous) functional $\mathcal{N}$ obtained from
equations (\ref{eq:buffer}), (\ref{eq:outrate}),
(\ref{eq:protocol}) and (\ref{eq:userflow2}). The notation
$\varkappa_t$ is here to emphasize that the evolution of the
network state depends on past state values \cite{Hale:93}. Note
that adjoining the flight-size expression is needed to obtain a
finite number of equilibrium points. Indeed, since the user flow
is computed from the derivative of the flight-size, the
equilibrium information is lost and can only be recovered from the
original expression of the flight-size. At equilibrium we indeed
have $\digamma_i^*=w_i^*=\RTT_i^*\phi_i^*$ where $\RTT_i^*$ and
$\phi_i^*$ are equilibrium values for RTT and the sending flow of
user $u_i$, respectively.

%
%
%
%

This model thus takes the form of a \emph{descriptor nonlinear
hybrid positive time-delay system with state-dependent and
constant delays} about which many theoretical questions are open:
well-posedness, existence of solutions, uniqueness of solutions,
stability of solutions, etc. Note also that in this paper, we have
not discussed about delay-derivative constraints whose violation
may lead to severe well-posedness problems \cite{Michiels:11}, such as nonuniqueness of solutions, existence of small-solutions, stopping solutions, etc. Some simple topologies have been considered in \cite{Briat:10}
where it is shown that delay-derivative may exceed one under
certain conditions. For the moment, it is unclear whether for
arbitrary topologies and under certain reasonable conditions, the
delays perceived by the users always have derivatives smaller than
one. This property is very suitable for analysis since many theoretical tools can only been applied when this condition is verified, e.g. Lyapunov-Krasovskii functionals \cite{GuKC:03} or certain integral quadratic constraints \cite{Kao:07}.

\subsection{Model Approximations}

The proposed framework includes explicit and seemingly exact
expressions for every quantity of interest, but this was not
historically the case. The sending rate model has always been a
missing link in past formulations where ad-hoc models were considered. It
has been recently shown in \cite{Jacobsson:08,Tang:10} that these
flow models are actually approximation of the W-ACK-clocking model, which is turn an approximation of the FS-ACK clocking model considered in this paper. We summarize these remarks below for completeness.

\subsubsection{Ratio flow model}

By making the approximation $\digamma_i(t)\simeq w_i(t)$ in the FS-ACK-clocking model to get the W-ACK clocking model, we obtain
\begin{equation}
  w_i(t)=\int_{\mathscr{B}_{C_i}(t)}^t\phi(u_i^+,s)ds
\end{equation}
and using the right-square rectangle rule we get the following expression for the sending rates
\begin{equation}
  \phi(u_i^+,t)\approx\dfrac{w_i(t)}{T_i+\tau(g^i(t))}
\end{equation}
which is very similar to the ratio flow model for instance considered in \cite{Misra:00, Vinnicombe:02, Paganini:03}.

\subsubsection{Joint flow model}

The joint flow model reuses the W-ACK-clocking model and by making a first order Taylor expansion on the implicit expression
\begin{equation}
  w_i(t)-\int_{t}^{\mathscr{F}_{C_i}(t)}\phi(u_i^+,s)ds=0
\end{equation}
we obtain
\begin{equation}
  w_i(t)+(T_i+\tau(t+T_i^f))(\dot{w}_i(t)-\phi(u_i^+,t))\approx0
\end{equation}
or equivalently
\begin{equation}
  \phi(u_i^+,t)\approx\dfrac{w_i(t)}{T_i+\tau(t+T_i^f)}+\dot{w}_i(t)
\end{equation}
which is exactly the joint flow model considered in \cite{Jacobsson:08,Moller:08,Jacobsson:08b,Jacobsson:09}. Neglecting the derivative term yields the usual ratio model \cite{Misra:00,Vinnicombe:02,Paganini:03}.

\subsubsection{Static model}\label{sec:static}

The static link model assumes that sending rates are proportional
to the derivative of congestion window sizes, making the relation
between these sizes and queuing delays static. This model can be
obtained by further approximating the ratio model or using
linearization and a (0,0) Pad\'{e} approximation. In Section
\ref{sec:hnct}, a more general proof for the static-link model is
provided and it suggests that the static model has a much wider
domain of validity, as experimentally emphasized in
\cite{Wang:05}.

\section{The Single-Buffer/Multiple-User Topology with Delay-based Protocols}\label{sec:SBMUtopology}

The purpose of this section is two-fold: exemplify the modeling technique on a simple topology and prove that when some conditions on the topology are met, the proposed model reduces to a model involving a static-link model \cite{Wang:05}. The proposed model hence allows to clarify the status of the static-link model \cite{Wang:05} by providing, for the first time, a mathematical proof for its domain of validity.

To this aim, a single-buffer/multiple-user
topology connected by lossless transmission channels is considered. The forward
and backward propagation delays of user $u_i$ are denoted by $T_i^f$
and $T_i^b$ respectively. We propose to use the following generic
model of any delay-based congestion control protocol as the user
model
\begin{equation}\label{eq:FAST}
  \begin{array}{lcl}
    \dot{z}_i(t)&=&\mathcal{P}_i(z_i(t),\tau(g^i(t)))\\
    w_i(t)&=&\mathcal{W}_i(z_i(t),\tau(g^i(t)))
  \end{array}
\end{equation}
where $z_i$, $w_i(t)$, $T_i=T_i^f+T_i^b$ and $g^i(t)=g(t-T_i^b)$ are the state of the protocol, the congestion window size, the propagation delay and the backward delay operator respectively. The functions $\mathcal{P}_i$ and $\mathcal{W}_i$ are defined as in (\ref{eq:protocol}).

\subsection{Multiple-User/Single-Buffer Topology Model}

The general model is given by (\ref{eq:generalmodel}) with (\ref{eq:FAST}) and

\begin{equation}\label{eq:general}
\small\begin{array}{rcl}
  \dot{\tau}(t)&=&\left\{\begin{array}{lcl}
    c^{-1}\sigma(t)+\delta(t)-1&& \mathrm{if\ }\mathcal{C}(t)\\
    0 && \mathrm{otherwise}
  \end{array}\right.\\
  \dot{\pi}_i(t)&=&\left\{\begin{array}{lcl}
      0 && \mathrm{if\ }\mathcal{T}_i(t)\\
      \dot{w}_i(t)+\phi(u_i^-,t) && \mathrm{otherwise}.
    \end{array}\right.\\
  \digamma_i(t)&=&\int_{g^i(t)-T_i^f}^t\phi(u_i^+,\theta)d\theta\\
  \phi(u_i^+,t)&=&\left\{\begin{array}{lcl}
      \dot{w}_i(t)+\phi(u_i^-,t) && \mathrm{if\ }\mathcal{T}_i(t)\\
      0&& \mathrm{otherwise}
    \end{array}\right.\\
  \phi(u_i^-,t)&=&\left\{\begin{array}{lcl}
    \frac{c\phi(u_i^+,g^i(t)-T_i^f)}{c\delta(g^i(t))+\sum_j\phi(u_j^+,g^i(t)-T_j^f)}&& \mathrm{if}\ \mathcal{C}(t)\\
    \phi(u_i^+,t-T_i^b-T_i^f) &&\mathrm{otherwise}
  \end{array}\right.\\
   \sigma(t)&=&\sum_{i=1}^{N}\phi(u_i^+,t-T_i^f)
\end{array}
\end{equation}
where $\delta(t)$ denotes the normalized cross-traffic ${\delta(t)\in[0,1)}$. The topology of this model is completely described by (\ref{eq:topology}) with the operators $\mathcal{B}$, $\mathcal{U}=\diag_{i=1}^N\{\mathcal{U}_i\}$, $\mathcal{R}_{ub}=\begin{bmatrix}
  \nabla_{T^b_1} & \ldots & \nabla_{T^b_N}
\end{bmatrix}^T$, $\mathcal{R}_{bb}=0$ and $\mathcal{R}_{bu}=\begin{bmatrix}
  \nabla_{T^f_1} & \ldots & \nabla_{T^f_N}
\end{bmatrix}$ where $\nabla_d$ is the constant delay operator with delay $d>0$.

%

\subsection{Homogeneous Delays and No-Cross Traffic - The Static-Link Model}\label{sec:hnct}

As stated in Section \ref{sec:static}, the static-flow model can
be obtained using various approximations, which suggest that the
static-model is only valid locally. This however contradicts the
results reported in \cite{Wang:05} where it is emphasized that the
static model may yield quite precise results over a wide domain. Note also that the static-link model has been invalidated in many scenarios, notably some involving very heterogeneous delays or cross-traffic, see e.g. \cite{Jacobsson:08} and examples of Section \ref{sec:validation}. In the following, we show that \emph{the static model can be exact
when some conditions on the network topology are met}.

Assuming that the propagation delays are homogeneous, i.e.
$T_i^f=T^f$, $T_i^b=T^b$, $i=1,\ldots,N$, and that there is no
cross-traffic, i.e. $\delta\equiv0$, the model in
(\ref{eq:general}) reduces to
\begin{equation}
\begin{array}{lcl}
  \dot{\tau}(t)&=&\left\{\begin{array}{lcl}
    c^{-1}\sigma(t)-1&& \mathrm{if\ }\mathcal{C}(t)\\
    0 && \mathrm{otherwise}
  \end{array}\right.\\
\dot{\pi}_i(t)&=&\left\{\begin{array}{lcl}
      0 && \mathrm{if\ }\mathcal{T}_i(t)\\
      \dot{w}_i(t)+\phi(u_i^-,t) && \mathrm{otherwise}.
    \end{array}\right.\\
\digamma_i(t)&=&\int_{g_b(t)-T^f}^t\phi(u_i^+,\theta)d\theta\\
\phi(u_i^+,t)&=&\left\{\begin{array}{lcl}
      \dot{w}_i(t)+\phi(u_i^-,t) && \mathrm{if\ }\mathcal{T}_i(t)\\
      0&& \mathrm{otherwise}
    \end{array}\right.\\
     \phi(u_i^-,t)&=&\left\{\begin{array}{lcl}
    \frac{c\phi(u_i^+,g_b(t)-T^f)}{\sum_j\phi(u_j^+,g_b(t)-T^f)}&& \mathrm{if}\ \mathcal{C}(t)\\
    \phi(u_i^+,t-T^b-T^f) &&\mathrm{otherwise}
  \end{array}\right.\\
\sigma(t)&=&\sum_{i=1}^{N}\phi(u_i^+,t-T^f)\\
      g_b(t)&=&g(t-T^b).
\end{array}
\end{equation}
Assuming further that the buffer is always congested (i.e. $\mathcal{C}(t)$ holds true for all $t$) and all users are active (i.e. the $\mathcal{T}_i(t)$'s are all true) we obtain
\begin{equation}
  \dot{\tau}(t)=c^{-1}\sum_i\dot{w}_i(t-T^f)
\end{equation}
after the substitution of sending flows in queue dynamics.
Integrating the above equation from $0$ to $t$ we obtain
\begin{equation}\label{eq:sl}
  \tau(t)=c^{-1}\sum_iw_i(t-T^f)-T
\end{equation}
where we assumed $\tau(0)=0$ and $w_i(0)=0$, $i=1,\ldots,N$. The additional constant term $-T$ can be determined such that the above equation satisfies the equilibrium equation $\sum_iw_i^*(T+\tau^*)=c$. Equation (\ref{eq:sl}) is \emph{exactly} the static model, showing then its exactness for the single-buffer/multiple-user topology with homogeneous propagation delays and no cross-traffic. This result might be generalizable to the case of chained buffers and multiple users. The exactness of the static model over more complex topologies is an open question. The case of single-buffer with constant cross-traffic may also be analyzable.

Thus, according to the model proposed in (\ref{eq:general}), the
static-link model is exact in the single-link topology whenever
\begin{itemize}
  \item the buffers are permanently congested, i.e. $\mathcal{C}(t)$ holds true for all $t\ge0$;
  \item the propagation delays are homogeneous, i.e. $T_i^f=T^f$, $T_i^b=T^b$, $i=1,\ldots,N$;
  \item the cross-traffic is absent, i.e. $\delta\equiv0$;
  \item the users are not in ACK-retaining mode, i.e. $\mathcal{T}_i(t)$ holds true for all $i=1,\ldots,N$.
\end{itemize}


Compared to the justification of this model in \cite{Tang:10}, the
proof developed above is much more insightful since no model
approximation is made, only assumptions on the network topology.
This shows that the static-link model has an application domain
which is much wider than the ratio-link and the joint-link models, when the conditions on the topology are met.
It is indeed valid in the nonlinear setting and it does not result
from any approximation, just assumptions on the topology.

This also shows that the proposed metamodel is able to provide
theoretical justifications of a simpler model. This supplies a way
for deriving proofs for validity domain of models.



\subsection{Homogeneous Delays and Cross-Traffic}

When cross-traffic is added to the problem, the overall picture
changes. The cross-traffic acts as a bandwidth limiter both in the
networking and control terminology. Indeed, a nonzero $\delta(t)$
reduces the maximal output capacity $c$, creating then sort of
`varying-output-capacity' $c(1-\delta(t))$ which reduces the
bandwidth perceived by users.

\begin{result}
In the congested mode, the queue model writes
\begin{equation}
\begin{array}{lcl}
  \dot{\tau}(t)&=&c^{-1}\sum_i\dot{w}_i(t-T^f)+\delta(t)-\varphi(t)\\
  \varphi(t)&=&\dfrac{c\delta(g_{bf}(t))}{c\delta(g_{bf}(t))+\sum_j\phi_j(u_j^+,g_{bf}(t)-T^f)}
\end{array}
\end{equation}
where $g_{bf}(t)=g(t-T^b-T^f)$ and $\varphi(t)$ is the output flow
at time $t$ corresponding to the cross-traffic. The term
$\varphi(t)$ is responsible for the bandwidth reduction.
\end{result}

\begin{proof}
  Since the buffer is always congested and the users are not in ACK-retaining mode, we have
  \begin{equation}
    \phi_i(u_i^+,t)=\dot{w}_i(t)+\phi_i(u_i^-,t)
  \end{equation}
  and
  \begin{equation}
    \phi_i(u_i^-,t)=\dfrac{c\phi_i(u_i^+,g_b(t)-T^f)}{c\delta(g_b(t))+\sum_j\phi_j(u_j^+,g_b(t)-T^f)}.
  \end{equation}
  Substituting the above expressions in the queue model and noting that $\sum_j\phi_j(u_j^-,t-T^f)+\varphi(t)=c$ yields the result.
\end{proof}
%

\section{Model Validation}\label{sec:validation}

We consider the scenarios given in \cite{Jacobsson:08,Tang:10} to validate the proposed model. The results obtained via NS-2 have been slightly shifted in time so that the congestion window variation times match. Unlike \cite{Tang:10}, the results are not shifted in amplitude and this causes small discrepancies. If, however, the NS-2 results were shifted vertically so that they match initial equilibrium values, then the curves would match almost perfectly. 

\subsection{Single-Buffer/Multiple-Users}

\begin{figure}[H]
  \centering
        \psfrag{b-}[c][c]     {\small{$b^-$}}
        \psfrag{b+}[c][c]     {\small{$b^+$}}
        \psfrag{u1-}[c][c]     {\small{$u_1^-$}}
        \psfrag{u1+}[c][c]     {\small{$u_1^+$}}
        \psfrag{u2-}[c][c]     {\small{$u_2^-$}}
        \psfrag{u2+}[c][c]     {\small{$u_2^+$}}
  \includegraphics[width=0.2\textwidth]{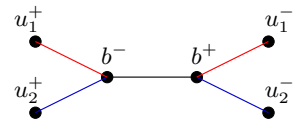}
  \caption{Topology of scenarios 1 \& 2}\label{fig:top1}
\end{figure}

\begin{figure}[H]
\begin{minipage}[b]{0.49\linewidth}
\centering
 \includegraphics[width=\textwidth]{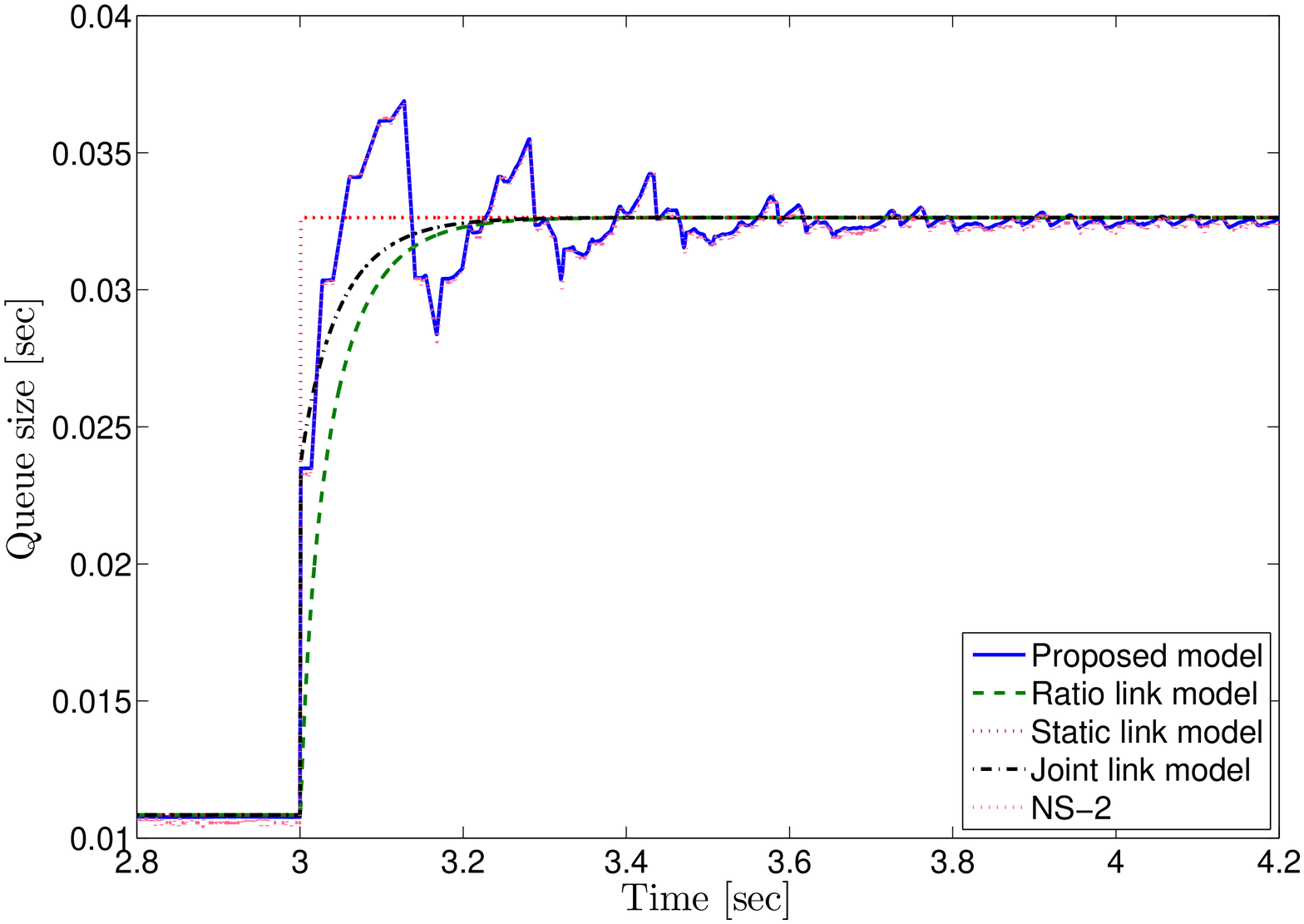}
\caption{Scenario 1: Queue size}\label{fig:ex1}
\end{minipage}
\hfill
\begin{minipage}[b]{0.49\linewidth}
\centering
 \includegraphics[width=\textwidth]{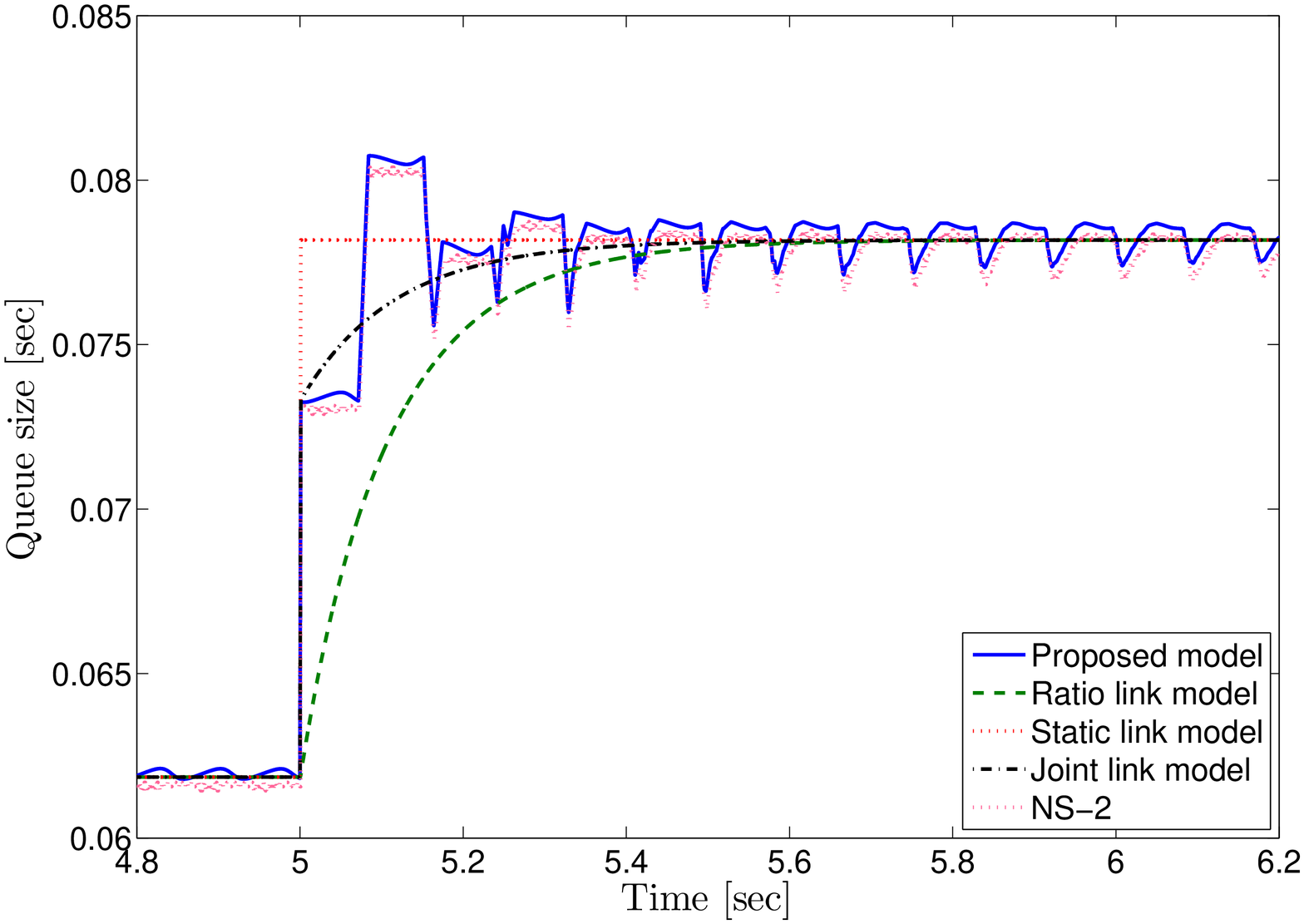}
\caption{Scenario 2: Queue size}\label{fig:ex2}
\end{minipage}
\end{figure}

In this section, we consider the interconnection of two users through a single resource, as depicted in Fig. \ref{fig:top1}. The bottleneck has capacity $c=100$Mb/s and the packet size including headers is 1590 bytes. The following scenarios from \cite{Jacobsson:08,Tang:10} are considered:
\begin{itemize}
  \item Scenario 1: the congestion window sizes are initially $w_1^0=50$ and $w_2^0=550$ packets, at 3s $w_1$ is increased to 150 packets. The propagation delays are $T_1=3.2$ms and $T_2=117$ms for users 1 and 2 respectively; see Fig. \ref{fig:ex1}.
  \item Scenario 2: the congestion window sizes are initially $w^0_1=210$ and $w_2^0=750$ packets, at 5s $w_1$ is increased to 300 packets. The propagation delays are $T_1=10$ms and $T_2=90$ms for users 1 and 2 respectively; see Fig. \ref{fig:ex2}.
\end{itemize}
We can see that our results fit well with the ones obtained by
packet level (NS-2) simulations. Yet, this is not the case for the
ratio-link, static-link, and the joint-link models when the
transient-state is considered. The obtained results by the
proposed model are identical to the results given in
\cite{Jacobsson:08,Tang:10} and obtained using the W-ACK-clocking model. This is expected since the W-ACK clocking is an approximation of the FS-ACK-clocking model defended in this paper. Note also that the approximation condition related to the window size increase is satisfied here.

\subsection{Multiple-Buffers/Multiple-Users}

\begin{figure}[H]
  \centering
        \psfrag{b1-}[c][c]     {\small{$b_1^-$}}
        \psfrag{b1+}[c][c]     {\small{$b_1^+$}}
        \psfrag{b2-}[c][c]     {\small{$b_2^-$}}
        \psfrag{b2+}[c][c]     {\small{$b_2^+$}}
        \psfrag{u1-}[c][c]     {\small{$u_1^-$}}
        \psfrag{u1+}[c][c]     {\small{$u_1^+$}}
        \psfrag{u2-}[c][c]     {\small{$u_2^-$}}
        \psfrag{u2+}[c][c]     {\small{$u_2^+$}}
        \psfrag{u3-}[c][c]     {\small{$u_3^-$}}
        \psfrag{u3+}[c][c]     {\small{$u_3^+$}}
        \psfrag{d-}[c][c]     {\small{$\delta^-$}}
        \psfrag{d+}[c][c]     {\small{$\delta^+$}}
  \includegraphics[width=0.35\textwidth]{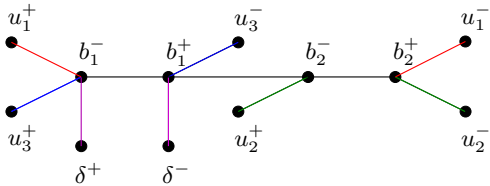}
  \caption{Topology of scenarios 3 to 6 ($\delta$ represents cross-traffic I/O nodes).}\label{fig:top2}
\end{figure}

\begin{figure}[H]
\begin{minipage}[b]{0.49\linewidth}
\centering
 \includegraphics[width=\textwidth]{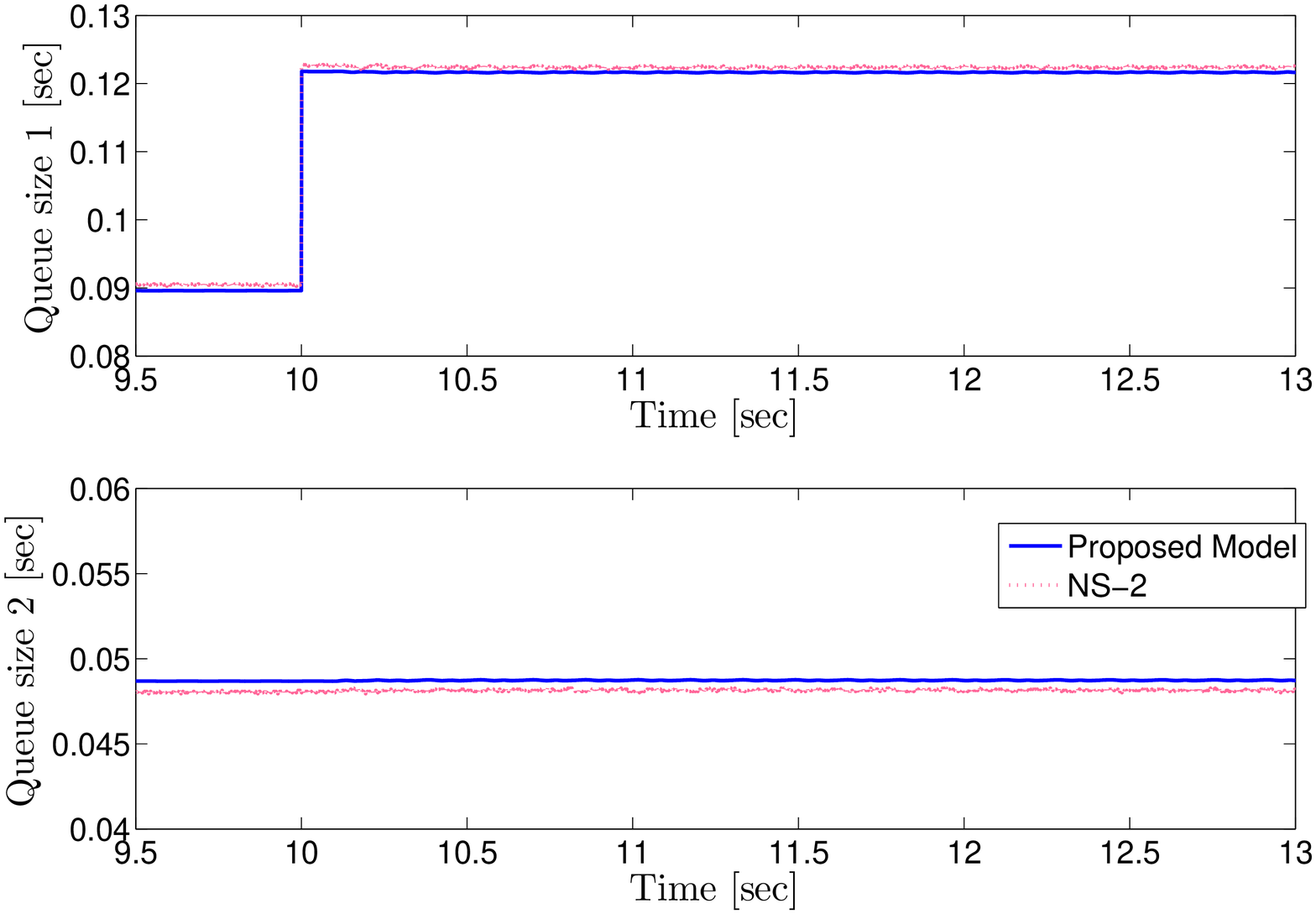}
\caption{Scenario 3: queue 1 (top) and queue 2 (bottom)}\label{fig:ex3}
\end{minipage}
\hfill
\begin{minipage}[b]{0.49\linewidth}
\centering
 \includegraphics[width=\textwidth]{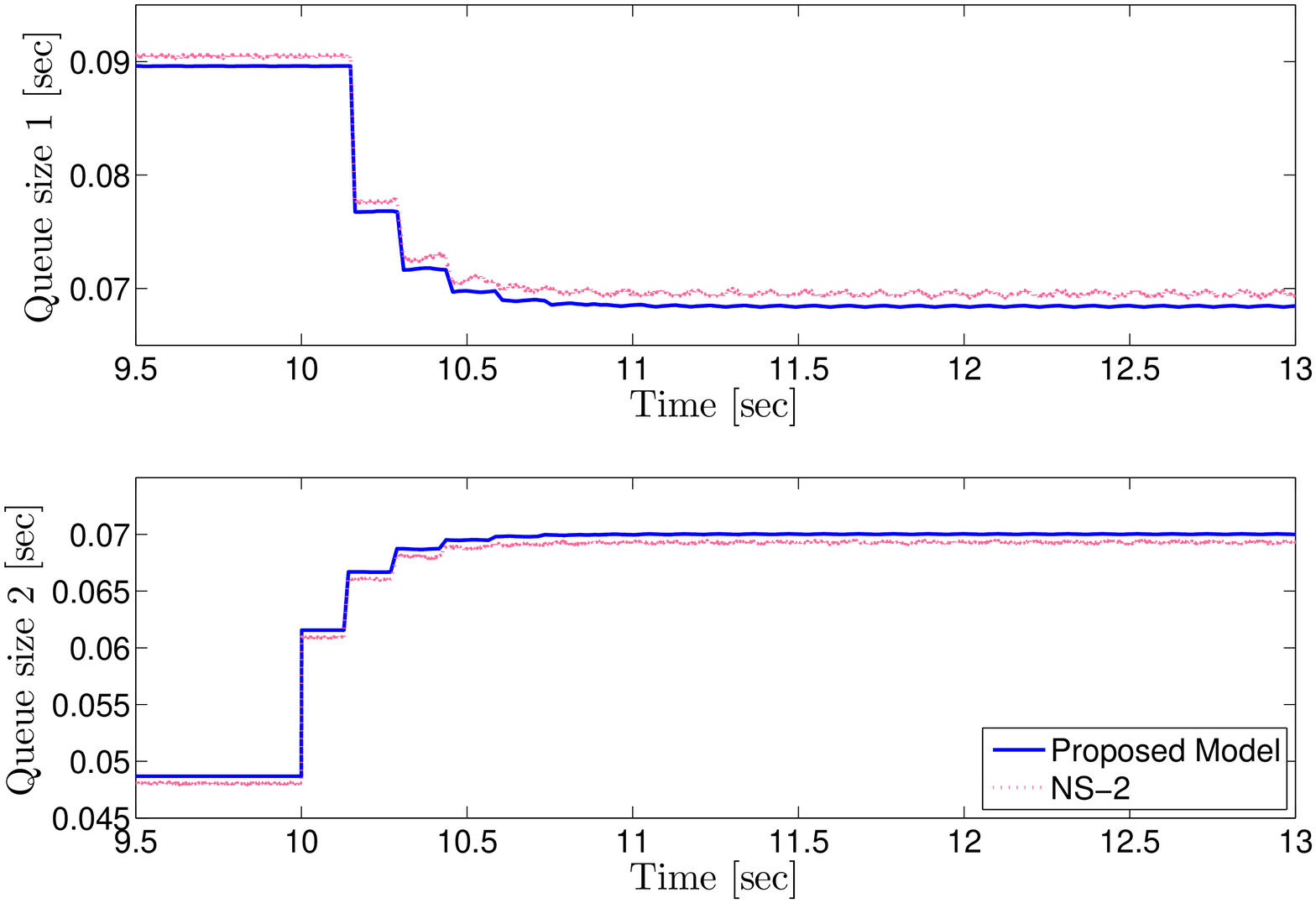}
\caption{Scenario 4: queue 1 (top) and queue 2 (bottom)}\label{fig:ex4}
\end{minipage}
\end{figure}

Here, we consider the case of two buffers interconnected in series (see Fig. \ref{fig:top2}) with capacities $c_1=72$Mb/s and $c_2=180$Mb/s. The packet size including headers is $1448$ bytes. The link propagation delays are $20$ms for link 1 and $40$ms for link 2. The total round-trip propagation delays are $T_1=120$ms, $T_2=80$ms and $T_3=40$ms for sources 1, 2 and 3 respectively. Initially, the congestion window sizes are $w_1^0=1600$ packets, $w_2^0=1200$ packets and $w_3^0=5$ packets. The following scenarios from \cite{Jacobsson:08,Tang:10} are considered:
\begin{itemize}
  \item Scenario 3: No cross-traffic and the congestion window $w_1$ is increased by 200 packets at 10s; see Fig. \ref{fig:ex3}.
  \item Scenario 4: No cross-traffic and the congestion window $w_2$ is increased by 200 packets at 10s; see Fig. \ref{fig:ex4}.
\end{itemize}
The proposed model is again able to capture the network behavior
well and it retrieves the previous results reported in
\cite{Jacobsson:08,Tang:10}. This is again due to the equivalence
between the ACK-clocking models in this case.

\begin{figure}[H]
\begin{minipage}[b]{0.49\linewidth}
\centering
 \includegraphics[width=\textwidth]{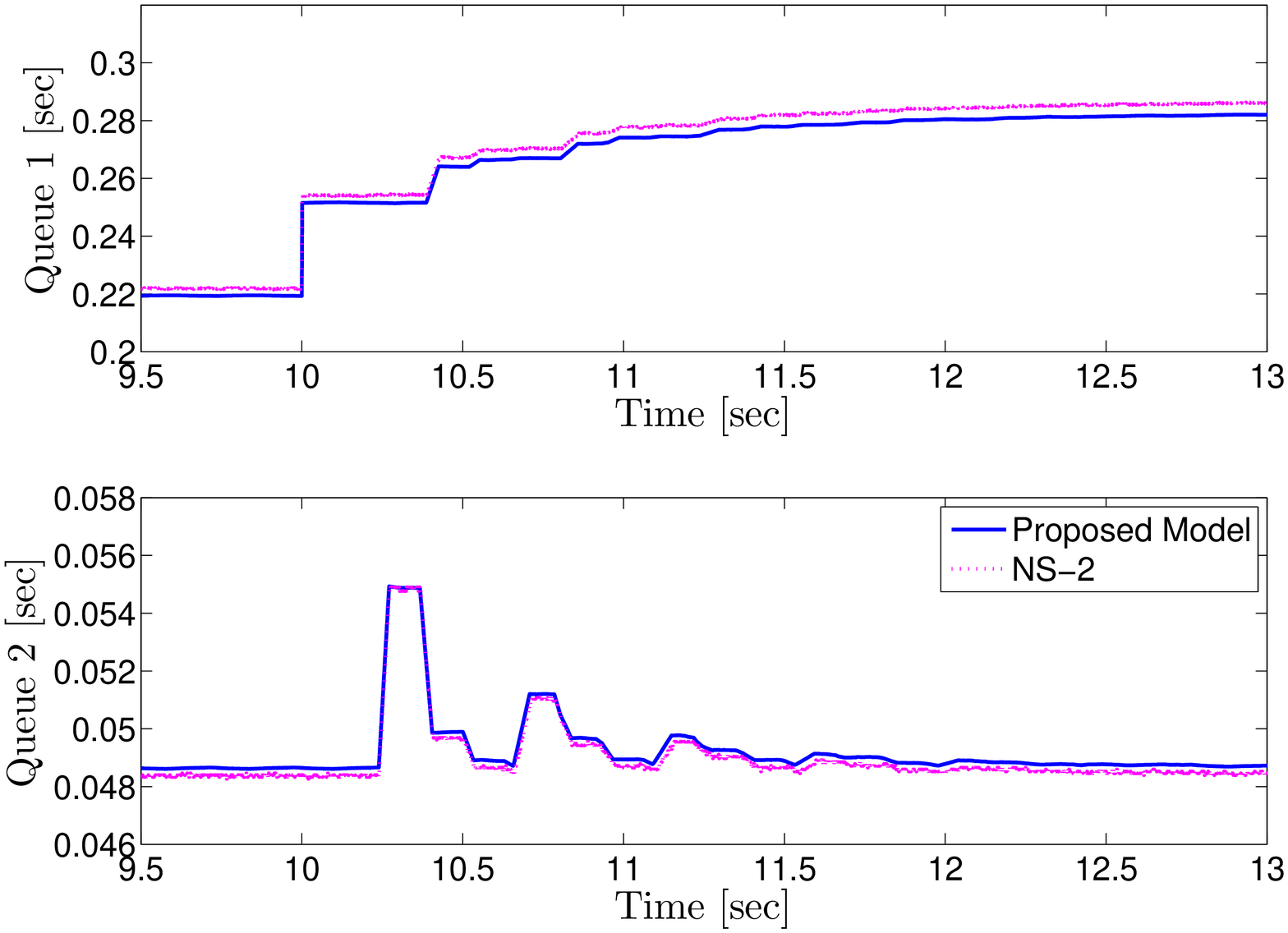}
\caption{Scenario 5: queue 1 (top) and queue 2 (bottom)}\label{fig:ex5}
\end{minipage}
\hfill
\begin{minipage}[b]{0.49\linewidth}
\centering
 \includegraphics[width=\textwidth]{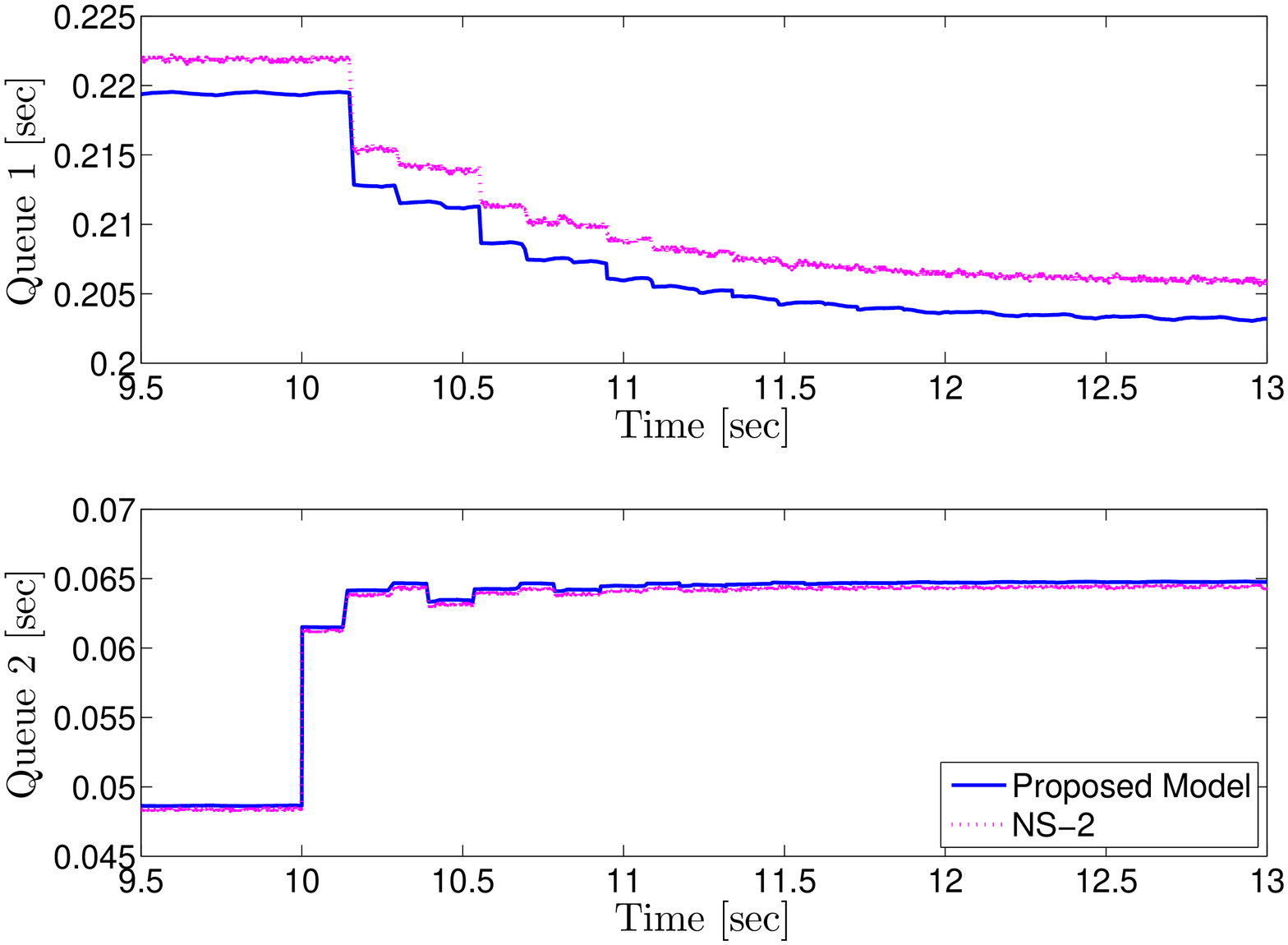}
\caption{Scenario 6: queue 1 (top) and queue 2 (bottom)}\label{fig:ex6}
\end{minipage}
\end{figure}

We introduce now a constant cross-traffic $x_{c1}=c_1/2$ on the
first link. Initially\footnote{This scenario is actually identical
to the one in \cite[Section III.B.3]{Tang:10}, the initial values
for congestion window sizes given in \cite{Tang:10} are incorrect.},
we set $w_1^0=1200$, $w_2^0=1600$, $w_3^0=5$ and we consider the
following scenarios:
\begin{itemize}
  \item Scenario 5: The congestion window $w_1$ is increased by 200 packets at 10s; see Fig. \ref{fig:ex5}.
  \item Scenario 6: The congestion window $w_2$ is increased by 200 packets at 10s; see Fig. \ref{fig:ex6}.
\end{itemize}

The obtained results are identical to the results obtained by the
the NS-2 simulations. Notice the reaction time between the moment
at which the congestion window size is increased and the moment at
which the second queue sees the flow variation. This illustrates
that the model captures well the communication path and the order
of elements (spatial and temporal topology). These characteristics
are not directly visible in the simulation results given in
\cite{Tang:10} since the curve steps seem to have been aligned on
the same temporal cursor.

Despite of results equivalence, the proposed metamodel enjoys interesting properties such as modularity and scalability, that the model reported in \cite{Jacobsson:08,Tang:10} lacks. This is an important improvement over previous models that were not able to cumulate accuracy, scalability, modularity and other interesting properties. This will be discussed in more detail in the next Section.

\subsection{Decreasing the Congestion Window Size}
The models proposed in \cite{Zhang:10,Tang:10} do not capture
sudden decreases in the congestion window size that would cause
the buffer to empty or become smaller than the actual flight-size,
that is, smaller than the number of packets in flight. The
proposed model does capture these phenomena since 1) the
FS-ACK-clocking model derived from the conservation law involves the
flight-size rather than the congestion window size, unlike in
\cite{Tang:10}; and 2) the user model implements an ACK-buffer to
count the number of packets to remove from the network before
starting to send again. Note that the derivation of the user model
including the ACK-buffer has been made possible due to the
availability of an explicit expression for the flow of
acknowledgments (\ref{eq:ackflow}). This makes it computable through
an explicit solution for the queuing delay and the buffer output
flows \cite{Briat:10,Briat:11k}. In \cite{Jacobsson:08}, the
decreasing of congestion window size is handled by adding a rate
limiter to constrain the (negative) slope of the queue size. This
rate limiter is however rather difficult to characterize due to
the time-varying nature of the lower-bound on the slope which
depends on the received rate of acknowledgment and the network
state, the former being unfortunately unavailable in the framework
of the thesis \cite{Jacobsson:08}.

%
%

\begin{figure}[H]
\begin{minipage}[b]{0.49\linewidth}
\centering
 \includegraphics[width=\textwidth]{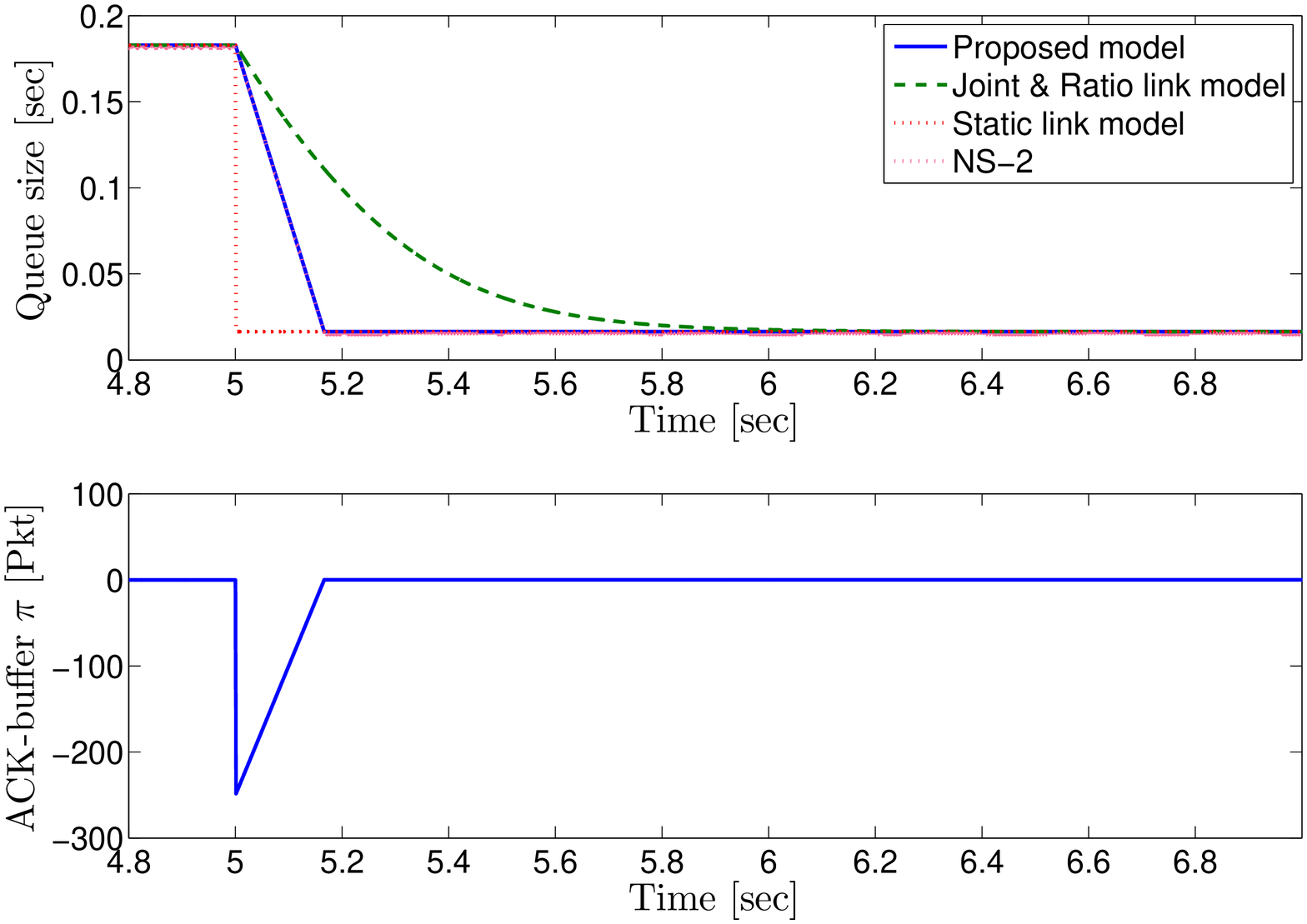}
\caption{Scenario 7: Queue size (top) ACK buffer (bottom)}\label{fig:ex_1k}
\end{minipage}
\hfill
\begin{minipage}[b]{0.49\linewidth}
\centering
  \includegraphics[width=\textwidth]{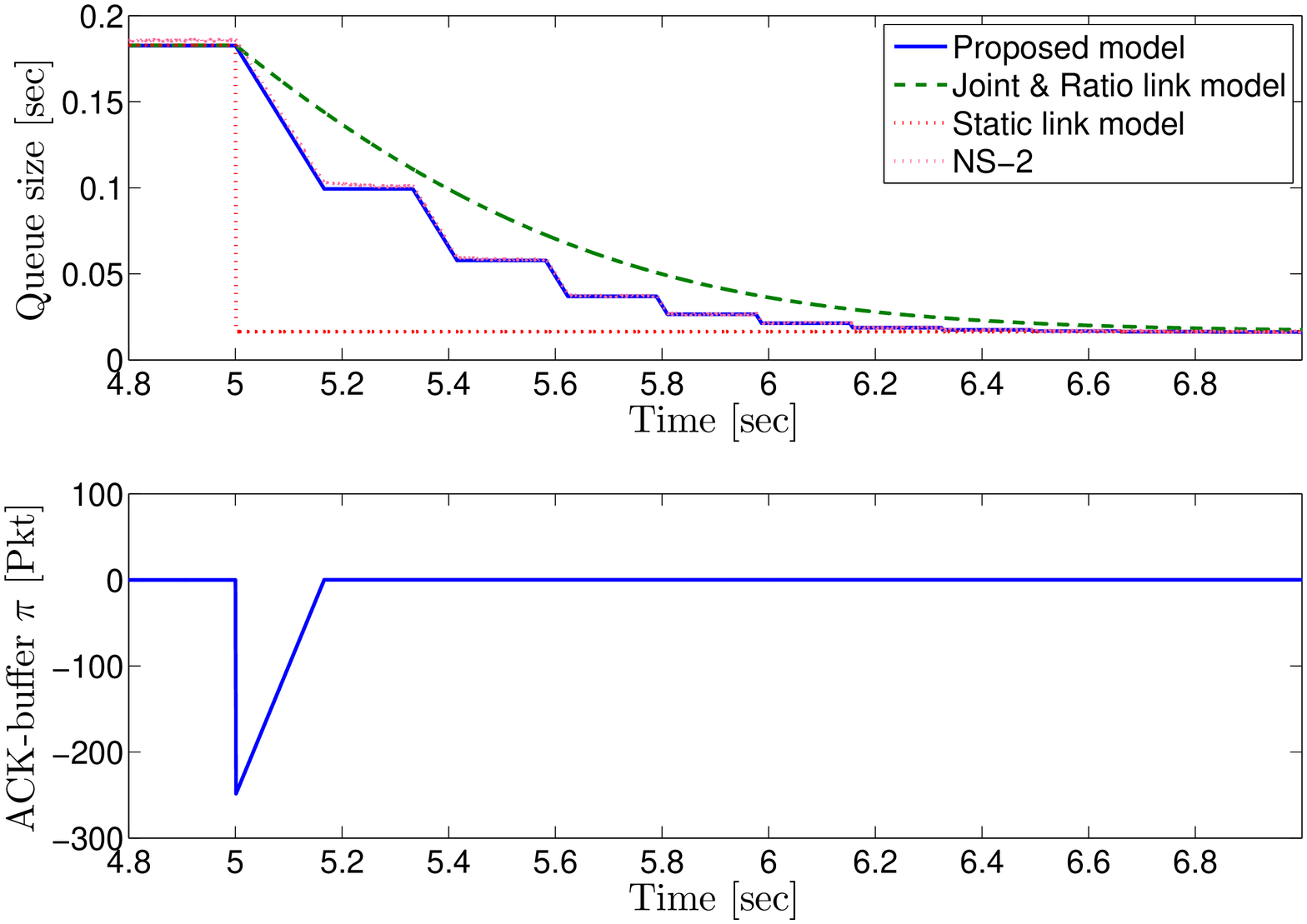}
\caption{Scenario 8: Queue size (top) ACK buffer (bottom)}\label{fig:ex_2k}
\end{minipage}
\end{figure}

Let us consider the single-user/single-buffer case where the total propagation delay is $T=150$ms, the packet size including headers is $1040$ bytes and the initial value of the congestion window size is $w^0=500$. A $t=5$ seconds, the congestion window size is halved. We consider the following scenarios
\begin{itemize}
  \item Scenario 7: $c=12.5$Mb/s and no cross-traffic; see Fig. \ref{fig:ex_1k}.
  \item Scenario 8: $c=25$Mb/s and half capacity used by cross-traffic; see Fig. \ref{fig:ex_2k}.
\end{itemize}
We can see that we obtain exactly the same results as NS-2 simulations (and the rate-limiter model reported in \cite{Jacobsson:08}). As desired, the ACK-buffer measures (counts) the number of packets to remove before starting to send again. 

\section{Related Work}\label{sec:related}

Despite of being unorthodox, it was to the authors' point of view
to discuss the related works after presenting the proposed
metamodel. Many congestion models have been proposed in the
literature with different formalisms: time-domains, protocols,
network topologies and other hypotheses. It is thus quite
difficult to compare them directly and give a thorough discussion.
Yet, we have identified some important characteristics which are
summarized in Tables \ref{tab:model}, \ref{tab:flow} and
\ref{tab:buffer}. Table \ref{tab:model} collects global
characteristics of a network model such as the \emph{time-domain}
property to indicate whether the model is in discrete (\emph{DT})
or continuous (\emph{CT}) time, or the \emph{modularity} property
to indicate whether it is modular or not. With \emph{delay} we
indicate that the model considers it either as constant
(\emph{Cst}), time-varying (\emph{TV}), state-dependent
(\emph{SD}), or asynchronous (\emph{Async}) discrete-time system
whose asynchrony (incorporating delays) depends on the state of
the system. \emph{Spatial} and \emph{temporal} topologies denote
whether the model share the same spatial and time topologies with
the actual network, respectively. \emph{Generic} indicates the
ability of the model to describe the network for different
protocols whereas \emph{rate model} denotes from which formula the
user sending rate is computed. The \emph{ratio} flow model \cite{Misra:00, Hespanha:01b, Vinnicombe:02, Farnqvist:02, Paganini:03} is
given by quotient of the congestion window size and the RTT. The \emph{static} flow model \cite{Wang:05} assumes that
the flow is proportional to the derivative of the congestion
window size and the \emph{joint} flow model \cite{Jacobsson:08,Moller:08,Jacobsson:08b,Jacobsson:09} assumes that the flow
is given by the sum of the ratio-flow model and the congestion
window size derivative.

Tables \ref{tab:flow} and \ref{tab:buffer} examine the important
subparts of the model, namely the user and buffer models,
respectively. It is also important to mention that none of them
has the structure of a metamodel having the precision of the
considered one. In Table \ref{tab:flow}, \emph{homogeneous} and
\emph{heterogeneous} delays denote how well a model captures the
case of homogeneous or heterogeneous propagation delays, whereas
\emph{cross} and \emph{bursty} traffic denote how well a model
captures the presence of cross traffic and bursty traffic \cite{Jiang:05b}, respectively.
\emph{Exact} indicates that the transient-state behavior of a
model matches well to the one observed in packet level
simulations. \emph{Too fast} and \emph{too slow} mean that a model
converges to steady-state faster or slower, respectively, compared
to what is observed in packet level simulations. \emph{Faster}
states a relativity faster convergence, yet not as much as
observed when it is \emph{too fast}. In Table \ref{tab:buffer},
\emph{output flows} indicates whether it is defined as
\emph{aggregate}, meaning all input flows are mixed, or
\emph{split}, where the buffer acts as a square MIMO system
mapping a given input flow to a given output flow. \emph{FIFO
characterization} denotes whether the model captures the actual
content of the queue and the FIFO behavior. With \emph{queuing
delay model} we indicate whether its effect on input flows are
clearly defined. \emph{Frequency filtering effect} denotes whether
the model of buffer have a filtering effect on the input flows to the output flows.

\begin{table*}
\centering
{\scriptsize  \begin{tabular}{lccccccccccc}
\toprule
\textbf{Network Models} & \cite{Johari:01} & \cite{Vinnicombe:02} & \cite{Misra:00} & \cite{Paganini:03b} & \cite{Wang:05} & \cite{Jacobsson:08} (DAE) & \cite{Tang:07} & \cite{Farnqvist:02} & \cite{Shorten:06} & \cite{Tang:10} & This paper\\
\midrule
\textbf{Characteristics}\\
  Time-domain & DT & CT & CT & CT & CT & DT & CT & Hybrid & DT & CT & Hybrid\\
  Delay       & Cst & Cst & TV/SD\footnotemark[1] & Cst & Cst & Async. & Cst & Cst & Async. & SD\footnotemark[2]. & SD\\
  Modular     & Almost\footnotemark[3] & Yes & No & Yes & Yes & No & Yes & Almost\footnotemark[4] & No & Almost\footnotemark[5] & Yes\\
  Spatial topology & Yes & No & Yes & No & Yes & Yes & Yes & No & Yes & Yes & Yes\\
  Temporal topology & Yes & No & Yes & No & Yes & Yes & Yes & Yes & Yes & Yes & Yes\\
  Generic   & Yes & No & No & No & No & No & No & No & Yes & Yes & Yes\\
  Rate model & No\footnotemark[6] & Ratio & Ratio & Ratio & Static & (FS)ACK\footnotemark[7] & Joint & Ratio & No\footnotemark[6] & (W)ACK\footnotemark[8] & (FS)ACK\footnotemark[9]\\
  \bottomrule
\end{tabular}}
\caption{Main characteristics of some existing models in the
literature.}\label{tab:model}
\end{table*}
\subsubsection{Table \ref{tab:model}}

{\footnotesize$^{1}$In this model, while the delay is denoted as a
function of time, it is implicitly defined from the state, without
further investigation. An exact delay characterization as a
function of the state has been obtained in \cite{Briat:10}.
$^{2}$The state-dependent delay is implicitly characterized by the
ACK-clocking formula and the RTT expression. $^{3}$The modularity
of the model could have been developed but was not. $^{4}$Modular
but not able to represent any topology, essentially due to lack of any
output flow model for buffers. $^{5}$This modeling technique can
be applied to any type of topology, however hand calculations,
which result in poor scalability, are necessary. $^{6}$The notion
of flow is not defined in a discrete-time model. $^{7}$The discrete-time model is derived
from a continuous-time model using the (FS)ACK-clocking model.
$^{8}$The flow is implicitly defined as the solution of the
(W)ACK-clocking model. $^{9}$The flow is computed by solving the
(FS)ACK-clocking model explicitly.}
\begin{table*}
\centering
{\scriptsize  \begin{tabular}{llccccc}
\toprule
& \multirow{2}{*}{\textbf{User sending flow models}} & \multirow{2}{*}{Static} & \multirow{2}{*}{Ratio} & \multirow{2}{*}{Joint} & \multirow{2}{*}{(W)ACK} & (FS)ACK+\\
 &&&&&&ACK-Buffer\footnotemark[10]\\
\midrule
\textbf{Congestion window} & \textbf{Characteristics}\\
\multirow{4}{*}{Increase} &  Homogeneous delays & Exact/Too fast\footnotemark[11] & Too slow\footnotemark[14] & Faster\footnotemark[12] & Exact & Exact\\
 & Heterogeneous delays & Too fast & Too slow\footnotemark[14] & Faster\footnotemark[12] & Exact & Exact\\
  &Cross-traffic & Too fast\footnotemark[13] & Too slow\footnotemark[14] & Faster\footnotemark[12] & Exact & Exact\\
  & Bursty-traffic & Exact/Too fast\footnotemark[15] & Too slow\footnotemark[14] & Faster\footnotemark[16] & Exact & Exact\\
\midrule
\multirow{4}{*}{Decrease} &  Homogeneous delays & Too fast\footnotemark[17] & Too slow\footnotemark[14] & Too slow\footnotemark[14] & Too fast\footnotemark[17] & Exact\\
 & Heterogeneous delays & Too fast\footnotemark[17] & Too slow\footnotemark[14] & Too slow\footnotemark[14] & Too fast\footnotemark[17] & Exact\\
 & Cross-traffic & Too fast\footnotemark[17] & Too slow\footnotemark[14] & Too slow\footnotemark[14] & Too fast\footnotemark[17] & Exact\\
\bottomrule
\end{tabular}}
\caption{Comparison of different user flow models with respect to
the accuracy of estimated queue sizes.}\label{tab:flow}
\end{table*}

\subsubsection{Table \ref{tab:flow}}

{\footnotesize $^{10}$This model is the one introduced in
\cite{Briat:11k} and also considered in the current paper.
$^{11}$In the single-buffer case the model is exact. It is unclear
whether this is also true for the multiple buffer case. In most
cases, the model is too fast. $^{12}$Accurate modeling of
discontinuity/high slope of the queue when burst of data arrives
at the queue input. $^{13}$The model is usually too fast. It might, however, be possible to refine it to account for constant cross-traffic
\cite{Moller:08}. $^{14}$Unable to characterize the high slope of
the queue due to the low-pass filtering effect of the resulting
queue model, see Section \ref{sec:compbuffer}. $^{15}$Exact in the case of homogeneous delays and single-buffer topology since the relation is static and it has an
infinite bandwidth; see Section \ref{sec:hnct}. Too fast otherwise. $^{16}$Larger bandwidth than the ratio model but still limited since the bandwidth reduces
as the queue size grows. $^{17}$These models may withdraw packets
from the network by injecting negative flows.}
\begin{table*}
\centering
{\scriptsize  \begin{tabular}{lcccc}
\toprule
\textbf{Buffer descriptions} & Aggregate model (\ref{eq:buffer})-(\ref{eq:outrate}) & Pseudo-queue model \cite{Farnqvist:02,Zhang:10}& Flow model \cite{Ohta:98,Liu:04,Briat:10}\\
\midrule
\textbf{Characteristics}\\
Output flows & Aggregate & Split & Split\\
FIFO characterization  & Lost\footnotemark[18] & No\footnotemark[19]& Yes\footnotemark[20]\\
Queuing delay model & Lost\footnotemark[18] & No\footnotemark[21] & Yes \footnotemark[22]\\
Frequency filtering effect  & All-pass/undef.\footnotemark[23]& All-pass/Low-pass\footnotemark[24]  & All-pass \footnotemark[25]\\
\bottomrule
\end{tabular}}
\caption{Comparisons between different buffer descriptions.}\label{tab:buffer}
\end{table*}
\subsubsection{Table \ref{tab:buffer}}

{\footnotesize $^{18}$The information is lost since the output
flows are aggregate. $^{19}$The output flows are insensitive to
content swapping. $^{20}$The output flows are sensitive to any
swap of information in the queue. $^{21}$The exact queuing delay
is actually difficult to assign to any output flow due to the
structure of the model for the output flows involving queues.
$^{22}$Undefined due to the aggregate formalism. $^{23}$All-pass
when the sum of the input-flows does not exceed the maximal output
capacity, otherwise the filtering effect is not really defined due
to the aggregate formalism. $^{24}$All-pass when the sum of the
input-flows does not exceed the maximal output capacity, otherwise
acts as a low-pass filter with bandwidth inversely proportional to
the queue size. $^{25}$This is due to the direct-feedthrough
structure of the model. The delay and nonlinear formula for the
output flows only have a compressing/expanding effect on time and
amplitude.}
%

\section{Conclusion}\label{sec:conclusion}

This paper presents a modular fluid-flow network congestion
control model to analyze communication networks with arbitrary
topology. This modular metamodel is introduced by mathematically modeling network elements such as queues, users, and
transmission channels, and network performance indicators such as
sending/acknowledgement rates and delays. It is
composed of building blocks that implement local mechanisms, some of which being ignored/unmodeled by some existing models in the literature. It is shown that the proposed model allows to recover existing models and brings a formal proof for their validity/invalidity as well as for their domain of validity. We present a
novel classification of the previously proposed models in the
literature and we show that the existing models are often not
capable of capturing the transient behavior of the network
precisely. Numerical results obtained from packet-level
simulations demonstrate the accuracy of the proposed model. We
plan to extend our work with modeling of data loss, such as packet
drops, and the time-out mechanism at the user level.

\section{Acknowledgment}

This work is dedicated to the memory of our colleague Prof. Ulf T. J\"{o}nsson, who unfortunately passed away before he could see it completed.

\bibliographystyle{IEEEtran}
%

\end{document}